\documentclass[journal]{IEEEtran}
\usepackage{amsmath} 
\usepackage{amssymb}
\usepackage{amsfonts}
\usepackage{amsthm}
\allowdisplaybreaks[4]
\usepackage{algorithmic}
\usepackage{algorithm}
\usepackage{gensymb}
\usepackage{array}
\usepackage{booktabs}
\usepackage[caption=false,font=normalsize,labelfont=sf,textfont=sf]{subfig}
\usepackage{textcomp}
\usepackage{color}
\usepackage{bm}
\usepackage{stfloats}
\usepackage{caption}
\usepackage{url}
\usepackage{verbatim}
\usepackage{graphicx}
\usepackage{diagbox}
\usepackage{epstopdf}
\usepackage{cite}

\newtheorem{lemma}{Lemma}
\newtheorem{prop}{Proposition}

\begin{document}

\title{Joint Spatial Registration and Resource Allocation for Transmissive RIS Enabled Cooperative ISCC Networks}

\author{Ziwei~Liu,~Wen~Chen,~Zhendong~Li,~and~Qiong~Wu
	

	\thanks{Z. Liu, and W. Chen are with the Department of Electronic Engineering, Shanghai Jiao Tong University, Shanghai 200240, China (e-mail: ziweiliu@sjtu.edu.cn; wenchen@sjtu.edu.cn).}
	\thanks{Z. Li is with the School of Information and Communication Engineering, Xi'an Jiaotong University, Xi'an 710049, China (e-mail: lizhendong@xjtu.edu.cn). }	
	\thanks{Q. Wu is with the School of Internet of Things Engineering, Jiangnan
	University, Wuxi 214122, China (e-mail: qiongwu@jiangnan.edu.cn). }

}

%

\maketitle

\begin{abstract}
In this paper, we propose a novel transmissive reconfigurable intelligent surface (TRIS) transceiver-driven cooperative integrated sensing, computing, and communication (ISCC) network to meet the requirement for a diverse network with low energy consumption. The cooperative base stations (BSs) are equipped with TRIS transceivers to accomplish sensing data acquisition, communication offloading, and computation in a time slot. In order to obtain higher cooperation gain, we utilize a signal-level spatial registration algorithm, which is realized by adjusting the beamwidth. Meanwhile, for more efficient offloading of the computational task, multistream communication is considered, and rank-$N$ constraints are introduced, which are handled using an iterative rank minimization (IRM) scheme. We construct an optimization problem with the objective function of minimizing the total energy consumption of the network to jointly optimize the beamforming matrix, time slot allocation, sensing data allocation and sensing beam scheduling variables. Due to the coupling of the variables, the proposed problem is a non-convex optimization problem, which we decouple and solve using a block coordinate descent (BCD) scheme. Finally, numerical simulation results confirm the superiority of the proposed scheme in improving the overall network performance and reducing the total energy consumption of the network.
\end{abstract}

\section{Introduction}
\IEEEPARstart{D}{riven} by 6G technology, the increasing abundance of wireless communication application scenarios, such as autonomous driving, industrial IoT, and smart cities, puts extreme demands on wireless service performance. Existing wireless networks are no longer able to meet the integrated demands for high-precision sensing, reliable communication, and efficient computation in these complex scenarios. To support promising applications, integrated sensing, computing, and communications (ISCC) has recently been recognized as an important enabler for completing the computation of raw data to enable accurate environmental sensing\cite{10574245}, improving the utilization of network resources and offering the possibility of meeting these demands, but the competition for communication and computational resources between the sensing tasks of ambient intelligence and the computational tasks of mobile devices is increasingly becoming a challenging issue\cite{9828481,10812728,10077112}.

Cooperative sensing technology can break through the limitations of single-node sensing and improve the accuracy, reliability and coverage of sensing through information sharing and cooperative work among multiple nodes. In complex wireless environments, multi-node collaboration can acquire environmental information more comprehensively and improve the detection, tracking and identification of targets, which is crucial for applications such as intelligent transportation and environmental monitoring\cite{10726912,10872780}.
The cooperative integrated sensing and communication (ISAC) system is investigated in \cite{10032141}, which simultaneously supports multiuser cooperative communication and direct radar target localization, providing additional cooperative gains for communication and sensing functions. In \cite{10769538}, it is found that deploying $N$ ISAC transceivers improves the average cooperative sensing performance across the network and reduces the Cram\'{e}r-Rao lower bound according to the $\ln ^{2}N$ scaling law, leading to a better S\&C performance tradeoff at the network level. Moreover, under challenging dynamic, high mobility scenario conditions, cooperative ISAC utilizing multiple base stations (BS) enables robust and fair UAV sensing performance, highlighting the potential of cooperative ISAC for reliable target detection and communication\cite{10824972}. To ensure effective multi-node cooperative sensing and communication, spatial registration is a key component. Accurate spatial registration can make the sensing information of each node match spatially accurately, avoiding errors in information fusion, and thus improving the overall sensing performance\cite{10273396}. An adjustable beam spatial registration algorithm is proposed to overlap each sensing region by adjusting the sensing beam width using signal-level fusion technique\cite{9800700}. Simulation results show that the proposed algorithm obtains higher spatial registration power gain, which results in higher power echo signals and hence better sensing capability. However, in practical applications, cooperative sensing faces challenges such as large data processing tasks and limited computational power, which can affect the efficiency and accuracy of cooperative sensing. 

Computing power also plays a central role in modern wireless systems. A large amount of sensory data needs to be processed and analyzed in real time to extract valuable information to support decision making. In \cite{10806723}, a novel ISCC system is investigated, aiming to efficiently serve different targets. The results show that the proposed algorithm can provide higher target estimation performance than state-of-the-art benchmarks, provided that communication and computational constraints are satisfied. Many sensing tasks have low latency requirements and are desired to be realized on terminals with limited computational power and energy supply. In \cite{10795233}, the effectiveness of utilizing the advanced computational power of mobile edge computing (MEC) servers and cloud servers to address terminal sensing tasks is investigated. The results show a significant reduction in the execution latency and energy consumption of the sensing task. Efficient computational resource allocation and management can ensure that system performance requirements are met while reducing energy consumption and computational costs. A sensing performance maximization problem is constructed in \cite{10159270} while guaranteeing the quality of service requirements of the task. Minimum resources are allocated to the computational tasks and the rest are used for the sensing tasks, and the results demonstrate the performance improvement resulting from the proposal. Multifunctional BSs performing downlink communication, target sensing, and edge computing are investigated in \cite{10613426}. Numerical results show that the proposed algorithm performs significantly better than full offloading and near-local offloading in the case of limited user power budget and abundant local energy resources. However, in complex environments, wireless signals are susceptible to multipath fading, occlusion, or interference, and the communication link is unstable, with degradation of data transmission rate and reliability. In high-density terminal scenarios, there is intense competition for wireless channel resources, which may trigger congestion and delay jitter. Meanwhile, in areas with insufficient edge network coverage, it is difficult for MEC to effectively serve local devices, and it is costly to deploy a large number of active relay devices and expand BSs. In addition, MEC nodes are limited by power supply capacity and need to reduce the overall energy consumption of the system.

Transmissive reconfigurable intelligent surface (TRIS), as an emerging technology, is able to significantly improve the signal propagation quality and enhance the communication and sensing capabilities through intelligent regulation of the wireless environment\cite{10680462,10740042,10242373}. It has the flexibility to change the amplitude, phase and polarization characteristics of the signal, enabling large-scale antenna transmission in a simple way with narrower beamwidths thus reducing interference\cite{10522473,10945425}. Furthermore, TRIS can be deployed as a relay within the coverage area of existing BS, eliminating coverage blind areas, expanding the MEC service range\cite{9982476}, ensuring stable access to edge devices, creating good conditions for cooperative sensing and efficient communication, and significantly reducing energy consumption and deployment costs due to its low-power characteristics\cite{11106465}, making it suitable for solving the challenges of ISCC networks. The joint beamforming design problem for RIS-enabled ISCC networks is considered in \cite{10816721} to improve the network performance under limited spectrum, energy resources, and complex interference management conditions, and the results show that the sensing error can be reduced by up to 92.2\%, while the computation rate and communication rate are improved by more than 29.5\% and 23.9\%, respectively. A simultaneous transmitting and reflecting RIS assisted ISCC framework for robotic networking is proposed in \cite{10643175}, the BS receives unloading signals from a decision-making robot and performs target robot sensing at the same time, and the results show that the proposed system improves the total sum computation rate. To overcome the performance limitations of a single ISAC BS, multiple access points are employed to perform target detection and multi-user communication simultaneously with the assistance of RIS. The results show the advantages of deploying RIS to assist in ISAC network\cite{10746496}. In \cite{10411853}, RIS is used to assist multiple BSs in jointly transmitting to multiple users while assisting multiple BSs in cooperative sensing to perform multiple target sensing. In \cite{10552127}, a RIS-assisted ISAC system is used to facilitate collaboration between multiple BSs and thereby increase the benefits of cooperation between communication and sensing capabilities. However, the existing work mainly focuses on the regulation of the channel by RIS and is mostly in the reflective mode of operation, without considering the transmissive case. In addition, cooperative schemes in ISCC networks are yet to be investigated.

Overall, the scenario requirements of future networks will be more diversified, involving communication, sensing and computation, which faces many problems such as resource allocation, cooperative transmission and energy consumption, and puts forward new requirements and challenges for future network design. However, the existing work seldom considers these factors comprehensively, and there is less research on the problem of multi-source information fusion for cooperative transmission, and the problem of spatial registration for multi-BS cooperation is not yet clear. The main contributions of this paper are as follows:
\begin{itemize}
	\item[$\bullet$] We establish a cooperative ISCC network architecture based on time slots. In this architecture, multiple ISCC devices cooperate to realize target sensing data acquisition and communicate with the access point (AP) to offload and compute the data. A spatial registration strategy is adopted for the cooperative sensing area to enhance the cooperation gain, and the ISCC devices divide the sensing data into local computation part and remote computation part, where the local part is computed at the ISCC devices and the remote computation part is offloaded to the MEC at the AP for computation. In this process, the ISCC device is equipped with a TRIS transceiver for energy consumption and beam modulation considerations. To the best of our knowledge, there are few studies that consider the spatial registration problem for cooperative transmission.
	\item[$\bullet$] On the basis of the architectural design, we propose a joint optimization algorithm for spatial registration based on adjustable beamwidths, which is based on the block coordinate descent (BCD) framework. Since the computational data offloading is performed using multi-stream transmission and the problem solving involves rank-$N$ constraints, we use an iterative rank minimization scheme to deal with it. Meanwhile, based on the signal-level fusion idea, the spatial registration of the cooperative sensing region is realized by adjusting the beamwidth of the ISCC device.
	\item[$\bullet$] The effectiveness of the proposed algorithm is confirmed by simulation verification. The performance in terms of improving the execute data is improved by 8.66\% compared to the traditional transceiver, the energy consumption is reduced by 35.36\% and the communication offloading rate is improved by 23.47\%. In addition, the proposed scheme is superior to other baseline schemes and offers a potential solution for future networks due to the introduction of TRIS.
\end{itemize}

\emph{Notations}: Scalars are denoted by lower-case letters, while vectors and matrices are represented by bold lower-case letters and bold upper-case letters, respectively. $|x|$ denotes the absolute value of a complex-valued scalar $x$, ${x^ * }$ denotes the conjugate operation, and $\left\| \bf x \right\|$ denotes the Euclidean norm of a complex-valued vector $\bf x$. For a square matrix ${\bf{X}}$, ${\rm{tr}}\left( {\bf{X}} \right)$, ${\rm{rank}}\left( {\bf{X}} \right)$, ${{\bf{X}}^H}$, ${\left[ {\bf{X}} \right]_{m,n}}$ and $\left\| {\bf{X}} \right\|$ denote its trace, rank, conjugate transpose, ${m,n}$-th entry, and matrix norm, respectively. ${\bf{X}} \succeq 0$ represents that ${\bf{X}}$ is a positive semidefinite matrix. In addition, ${\mathbb{C}^{M \times N}}$ denotes the space of ${M \times N}$ complex matrices. $j$ denotes the imaginary element, i.e., $j^2 = -1$. The distribution of a circularly symmetric complex Gaussian (CSCG) random vector with mean $\mu $ and variance $\sigma^2$ is denoted by ${\cal C}{\cal N}\left( {{\mu},\sigma^2} \right)$ and $ \sim $ stands for ‘distributed as’. $\mathbb{E}\left(\cdot\right)$ represents the expectation of random
variables. ${\bf{A}} \otimes {\bf{B}}$ represents the Kronecker product of matrices ${\bf{A}}$ and ${\bf{B}}$. ${\bf{A}} \circ {\bf{B}}$ denotes the Hadamard product of matrices ${\bf{A}}$ and ${\bf{B}}$.

\section{System Model and Problem Formulation}
In this section, we first introduce the characteristics of TRIS and its necessity, followed by giving the system model and focusing on the mechanism of ISCC. Then explore signal and data fusion from multiple sources. Finally, describe the problem formulation.
\subsection{The characteristics and necessity of TRIS}
\begin{figure}[!htbp]
	\centerline{\includegraphics[width=5.5cm]{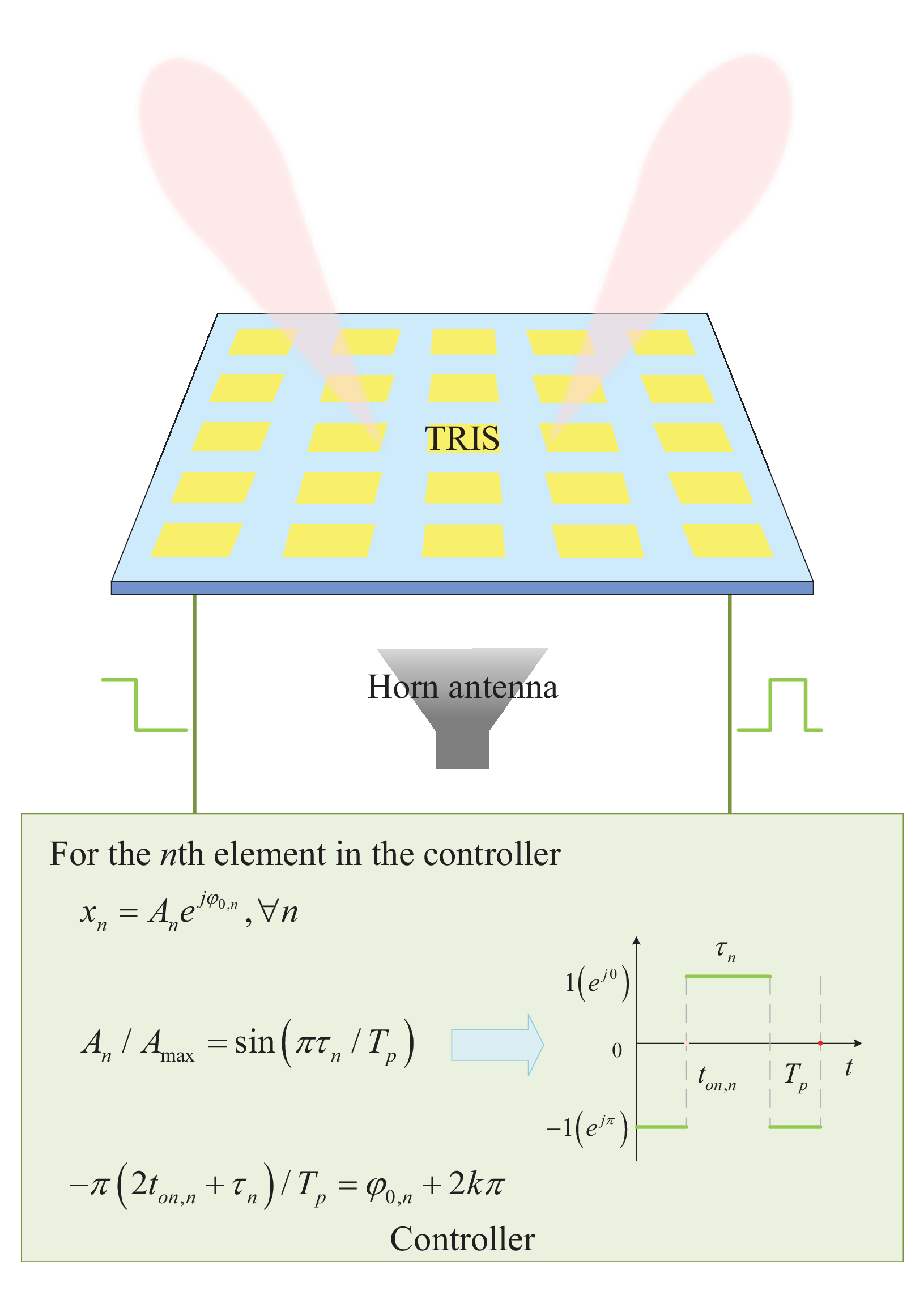}}
	\caption{The principles and compositions of TRIS transceiver.}\label{tris}
\end{figure}

ISCC devices can be roadside devices, unmanned aerial vehicles or vehicle side sensing devices with sensing, communication and computation capabilities, however, due to the limited capability of a single ISCC device, multiple ISCC devices collaborate to bring greater gains but face problems of high network power consumption and multi-target beam regulation. The TRIS transceiver, as a low-power consumption and low-cost device, bring a solution to this problem and consist of a TRIS with $N_t=N_r\times N_c$ reconfigurable transmissive elements arranged in a uniform planar array (UPA) pattern, a horn antenna, and a controller, with the principle and the composition illustrated in Fig. \ref{tris}. 

The TRIS transceiver avoids the need for the dense radiation array, power division network, coupling calibration network and blind connectors of the traditional multi-antenna transceiver, only a single RF link is required, so TRIS is simpler to implement, consumes less power, and is ready to meet the needs of future networks. Meanwhile, TRIS is easier to realize large-scale array with higher spatial diversity gain as well as more flexible beam regulation capability, and combined with time modulation array\footnote{As the core technology of TRIS, TMA employs Fourier series to modulate the amplitude and phase of the transmitted signal onto positive first harmonics according to the rules shown in Fig. \ref{tris}. Corresponding parameters are mapped to time-series signals with adjustable zero-state duration $\tau_{n}$ and zero-state onset time $t_{on,n}$. Subsequently, the time-series signals are loaded onto each element via control lines. Direct digital modulation is realized after the carrier signal of the horn antenna transmitting a single tone reaches the surface and the code element is recovered on the user side by extracting the harmonic components, where $T_p$ denotes the code element time. } (TMA) can support multi-stream transmission, which is more suitable for multi-target and collaboration scenarios. Furthermore, TRIS facilitates multi-beam and interference suppression more readily, effectively enhancing the capacity of edge links and providing a viable solution for MEC. Given the aforementioned advantages of TRIS deployment, energy efficiency, and multi-stream transmission, it holds significant application value in future networks.  However, it faces challenges such as power constraint handling, received signal processing, and unclear multi-station collaboration mechanisms, which will be explored in this paper. Regarding the impact of hardware parameters and channel imperfections on performance, a detailed study was conducted in \cite{11106465}.

\subsection{Network Model and Mechanisms of ISCC}
As shown in Fig. 2, we consider an ISCC network, which consists of an access point (AP) with MEC, $K$ ISCC devices, and $M$ targets. The AP has $N$ antennas arranged in a uniform linear array (ULA) pattern and is connected to the MEC by fiber optics with negligible latency. The antennas or elements of both AP and TRIS transceivers are spaced at half-wave lengths. Targets are considered point targets. Devices equipped with TRIS sense multiple targets through cooperative and scheduling mechanisms. The sensing data can be utilized for target velocity and range measurement, target classification and identification, as well as high-resolution imaging and reconstruction. Computational tasks involved in these operations are divided into local and remote computation. Communication serves to offload computational data and transmit collaborative scheduling information throughout this process.
\begin{figure}[!htbp]
		\centerline{\includegraphics[width=9.5cm]{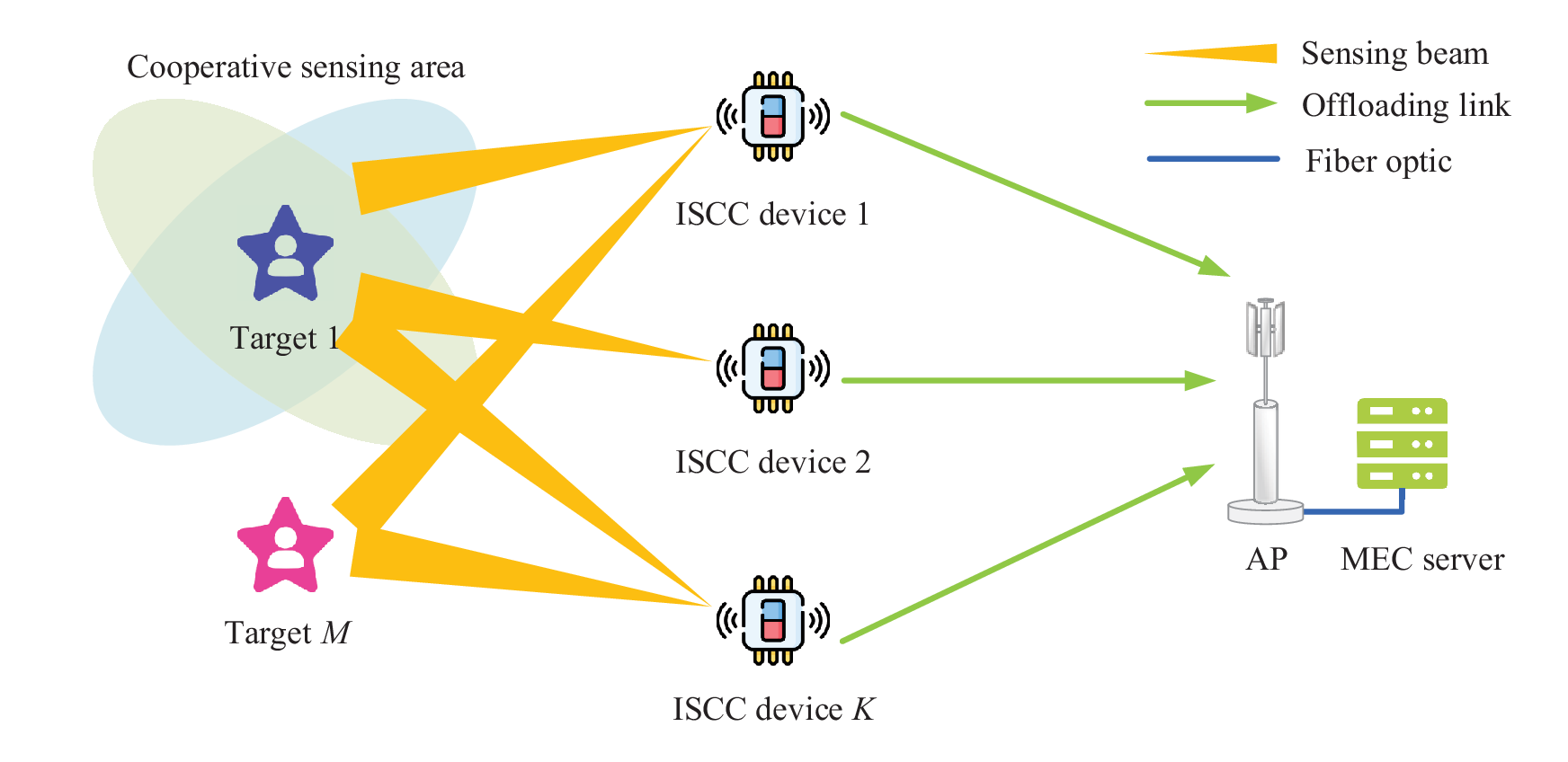}}
	\caption{ISCC networks.}
\end{figure}

The mechanism of ISCC network can be described as follows, we divide the total time duration $T$ into three parts, firstly, in the first time slot $t_{k}^{\rm I}$ the ISCC devices utilizes the rate-splitting signal to share and synchronize information with AP in preparation for cooperation, while simultaneously obtaining sensing data $D_{k}$. Then, the ISCC devices split the sensing data into a local computation portion $d_{k}^{l}$ and a remote computation portion $d_{k}^{r}$. Offload the remote computation portion $d_{k}^{r}$ to the AP via the private stream within the second time slot $t_{k}^{\rm II}$ and perform spatial registration of the cooperative sensing area in preparation for the transmission of the next frame. Meanwhile, the ISCC devices perform calculations on the local data $d_{k}^{l}$ for the remaining time $T-t_{k}^{\rm I}$. Finally, the AP performs remote computation tasks in the three time slot $t_{k}^{\rm III}$. This process is illustrated in Fig. 3 and the sensing data and time slots need to satisfy the following constraints
\begin{equation}
	d_{k}^{l}+d_{k}^{r}=D_{k},\forall k,
\end{equation}
and
\begin{equation}
	t_{k}^{\rm I}+t_{k}^{\rm II}+t_{k}^{\rm III}\le T,\forall k.
\end{equation}
\vspace{-10 mm} 
\begin{figure}[!htbp]
	\centerline{\includegraphics[width=9.5cm]{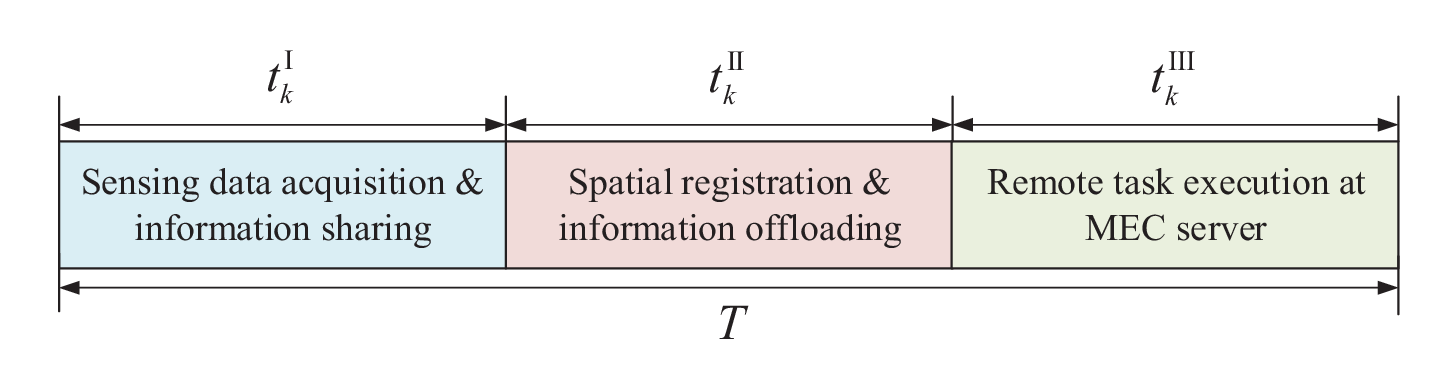}}
	\caption{Mechanisms of ISCC.}
\end{figure}
\vspace{-5 mm}
\subsection{Channel Model} 
\subsubsection{Radar Channel}
Based on the structure of TRIS, the steering vector of its downlink sensing link between the $k$th ISCC device and the $m$th target can be expressed as
\begin{equation}
\begin{split}
	{\bf a}\left(\theta_{k,m},\varphi_{k,m}\right)&={\left[ {{e^{ - j\pi {{\sin {\theta _{k,m}}\cos {\varphi _{k,m}}}}{{\bf{n}}_r}}}} \right]}\\ &~~~~~~~\otimes {\left[ {{e^{ - j\pi{{\sin {\theta _{k,m}}\sin {\varphi _{k,m}}}}{{\bf{n}}_c}}}} \right]},\forall k,m,
\end{split}
\end{equation}
where ${{\bf{n}}_r} = \left[ {0,1, \cdots ,{N_r} - 1} \right]^T$, ${{\bf{n}}_c} = \left[ {0,1, \cdots ,{N_c} - 1} \right]^T$, and $\left( {{n_r},{n_c}} \right)$ denotes the element position index of the TRIS.  ${\varphi _{k,m}}$ and ${\theta _{k,m}}$ denote the azimuth and pitch angles between the $k$th ISCC device and the $m$th target, respectively. The radar sensing mode employs a multimodal approach, specifically multiple transmit and receive channels. Since the TRIS transceiver has only a single antenna for reception, its received channel gain can be expressed as
\begin{equation}
	b_{i,m}=e^{-j2\pi d_{i,m}/\lambda_c},\forall i,m,
\end{equation}
where ${\lambda _c}$ represents the carrier wavelength, and ${d_{i,m}}$ represents the distance between the $i$th ISCC device and the $m$th target.
Thus the radar channel ${{\bf z}_{k,m,i}}\in\mathbb{C}^{1\times N_t}$ can be expressed as
\begin{equation}
	{{\bf z}_{k,m,i}}	= \alpha_{k,m}b_{i,m}{\bf a}^H\left(\theta_{k,m},\varphi_{k,m}\right),\forall k,m,i,
\end{equation}
where $\alpha_{k,m,i}\sim {\cal C}{\cal N}\left( {0,\sigma _{{\alpha _{k,m,i}}}^2} \right)$ denotes the complex reflection coefficient. $\sigma _{{\alpha _{k,m,i}}}^2 = {S_{RCS}} {\lambda_c}^2 / {(4\pi)^3d_{k,m}^2d_{i,m}^2}$, and ${S_{RCS}}$ denotes the RCS of target.
\subsubsection{Communication Channel}
For the communication channel ${\bf H}_k\in\mathbb{C}^{N\times N_t}$ between the $k$th ISCC device and the AP, which can be expressed as
\begin{equation}
	{\bf H}_k= \frac{\lambda_c}{4\pi d_k}\left( {\sqrt {\frac{{{\kappa}}}{{{\kappa} + 1}}} {{\bf{H}} _{k}^{LoS}} + \sqrt {\frac{1}{{{\kappa} + 1}}} {{\bf{H}}_{k}^{NLoS}}} \right),\forall k,
\end{equation}
where ${\bf{H}}_{k}^{NLoS}\sim{\cal CN}\left(0,{\bf I}_{N\times N_t}\right)$, $d_k$ denotes the distance between the $k$th ISCC device and the AP, and  ${\bf{H}}_{k}^{LoS}\in\mathbb{C}^{N\times N_t}$ can be expressed as
\begin{equation}
				\setlength{\abovedisplayskip}{3pt}
	\setlength{\belowdisplayskip}{3pt}
	{\bf{H}}_{k}^{LoS}={\bf e}_r\left(\theta_{k}\right){\bf e}_t^H\left(\theta_{k},\varphi_{k}\right),\forall k,
\end{equation}
where ${\bf e}_t\left(\theta_{k},\varphi_{k}\right)\in\mathbb{C}^{N_t\times 1}$ and ${\bf e}_r\left(\theta_{k}\right)\in\mathbb{C}^{N\times 1}$ denote the transmit and receive steering vectors, respectively. ${\varphi _{k}}$ and ${\theta _{k}}$ denote the azimuth and pitch angles between the $k$th ISCC device and the AP, respectively. Since the ISCC device employs the TRIS transceiver in UPA arrangement, its transmit steering vector can be expressed as
\begin{equation}
				\setlength{\abovedisplayskip}{3pt}
	\setlength{\belowdisplayskip}{3pt}
	{\bf e}_t\left(\theta_{k},\varphi_{k}\right)={\bf a}\left(\theta_{k},\varphi_{k}\right),\forall k.
\end{equation}
The antenna of the AP adopts the ULA arrangement, its receive steering vector can be expressed as
\begin{equation}
				\setlength{\abovedisplayskip}{3pt}
	\setlength{\belowdisplayskip}{3pt}
	{\bf e}_r\left(\theta_{k}\right)={\left[1, {{e^{ - j\pi {{\sin {\theta _{k}}}}}}},\cdots,{{e^{ - j\pi {{\sin {\theta _{k}}}}\left(N-1\right)}}} \right]^T},\forall k.
\end{equation}
The channels described above are assumed to be known and can be obtained by channel estimation methods\cite{8835503,9839429,9732214}.
\subsection{Signal Model}
In the first time slot $t_{k}^{\rm I}$, we utilize the deterministic radar signal for sensing and utilize the random signal for communicating with AP, and the signal ${\bf x}_k^{\rm I}\in\mathbb{C}^{N_t\times 1}$ sent by the ISCC devices  can be expressed as
\begin{equation}
	{\bf x}_k^{\rm I}={\bf W}_k^{\rm I}{\bf s}_k={\bf W}_{s,k}^{\rm I}{\bf s}_{s,k}+{\bf W}_{c,k}^{\rm I}{\bf s}_{c,k},\forall k,\label{I}
\end{equation}
where ${\bf W}_{k}^{\rm I}=\left[{\bf W}_{s,k}^{\rm I},{\bf W}_{c,k}^{\rm I}\right]\in\mathbb{C}^{N_t\times (N_t+G)}$ denotes the beamforming matrix in the first time slot. ${\bf W}_{s,k}^{\rm I}\in\mathbb{C}^{N_t\times N_t}$ and ${\bf W}_{c,k}^{\rm I}\in\mathbb{C}^{N_t\times G}$ denote the sensing stream beamforming matrix and communication stream beamforming matrix, respectively. $G\le\min\left\{N_t,N\right\}$ denotes the number of supported multistreams. To facilitate analysis, we assume that the symbols for sensing and communications the are independently and identically distributed\footnote{Within the first time slot, the sensing information and the communication information do not have an inclusion relationship and are therefore independent. Within the second time slot, the offloading information and the sensing information are also independent since the offloading information is derived from the sensing data of the first time slot.}. Therefore there are data ${\bf s}_k=\left[{\bf s}_{s,k},{\bf s}_{c,k}\right]=\left[s^s_1,s^s_2,\cdots,s^s_{N_t},s^c_1,s^c_2,\cdots,s^c_G\right]^T\in\mathbb{C}^{N_t+G}$  and $\mathbb{E}\left({\bf s}_k{\bf s}_k^H\right)={\bf I}_{N_t+G}$.

The power constraint of a traditional multi-antenna transceiver is reflected in the columns of the precoding matrix. The sum of the norms of the columns of the precoding matrix is required not to exceed the maximum transmit power. In this paper, a TRIS transceiver is used which employs the TMA as a modulation method and utilizes the generated temporal signals to control the phases of the TRIS elements, and matrix multiplication is involved in the process, so that the following requirements are imposed on the rows of the precoding matrix
\begin{equation}
	\left[{\overline{\bf W}}_{k}^{\rm I}\right]_{nn}\le P_t,\forall k,n,
\end{equation}
where  $P_t$ denotes the available maximum power of each TRIS element, ${\overline{\bf W}}_k^{\rm I}={\overline{\bf W}}_{s,k}^{\rm I}+{\overline{\bf W}}_{c,k}^{\rm I}$, ${\overline{\bf W}}_{s,k}^{\rm I}={\bf W}_{s,k}^{\rm I}\left({\bf W}_{s,k}^{\rm I}\right)^H$ and ${\overline{\bf W}}_{c,k}^{\rm I}={\bf W}_{c,k}^{\rm I}\left({\bf W}_{c,k}^{\rm I}\right)^H$. 

In this paper, a cooperative solution is deployed, so the ISCC devices are required to share information with the AP during the first time slot, including the location, communication protocol, serial number, IP address, control information and synchronization information, etc. The signal received by the AP in the first time slot can be expressed as
\begin{equation}
	\setlength{\abovedisplayskip}{3pt}
	\setlength{\belowdisplayskip}{3pt}
	{\bf y}^{\rm I}=\sum_{k=1}^{K}{\bf H}_k{\bf x}_k^{\rm I}+{\bf n},
\end{equation}
where ${\bf n}\sim{\cal CN}\left(0,\sigma_{n_k}^2{\bf I}_N\right)$ denotes the Gaussian noise vector at the AP. Then the achievable rate at the AP with respect to the $k$th ISCC device can be expressed as
\begin{equation}
				\setlength{\abovedisplayskip}{3pt}
	\setlength{\belowdisplayskip}{3pt}
	r_k^{\rm I}=\log_2{\rm det}\left({\bf I}_G+{\bf H}_k{\overline{\bf W}}_{c,k}^{\rm I}{\bf H}_k^H\left({\bf R}_k^{\rm I}\right)^{-1}\right),\forall k,
\end{equation}
where ${\bf R}_k^{\rm I}\!=\!\sigma_{n_k}^2{\bf I}_N+\sum\nolimits_{i \neq k}^K{\bf H}_k{\overline{\bf W}}_{c,i}^{\rm I}{\bf H}_k^H+\sum\nolimits_{i = 1}^K{\bf H}_k{\overline{\bf W}}_{s,i}^{\rm I}{\bf H}_k^H$.

TRIS supports multi-stream transmission and can separate echo signals through signal processing, so that different targets can be distinguished in terms of distance, angle and speed. We utilize the sensing scheduling variable $\rho_{k,m}\in\left\{0,1\right\}$ to indicate whether the $k$th ISCC device senses the $m$th target, so after reflection from the target, the received echo signal of the $k$th ISCC device can be expressed as
\begin{equation}
					\setlength{\abovedisplayskip}{3pt}
	\setlength{\belowdisplayskip}{3pt}
	{y}_k=\sum_{i=1}^{K}\sum_{m=1}^{M}\rho_{k,m}{\bf z}_{k,m,i}{\bf x}_i^{\rm I}+n_k,\forall k,
\end{equation}
where $n_k\sim {\cal C}{\cal N}\left( {0,{\sigma _{{k}}^2}} \right)$ denotes the Gaussian white noise at the ISCC device. Then the signal to interference plus noise ratio (SINR) of the $k$th ISCC device's echo about the $m$th target can be expressed as follow. 
\begin{equation}
					\setlength{\abovedisplayskip}{3pt}
	\setlength{\belowdisplayskip}{3pt}
	{SINR}^{echo}_{k,m}=\frac{\sum\limits_{i=1}^K\left\|\rho_{k,m}{\bf z}_{k,m,i}{\bf W}_{i}^{\rm I}\right\|^2}{\sum\limits_{j\neq m}^{M}\sum\limits_{i=1}^{K}\left\|\rho_{k,j}{\bf z}_{k,j,i}{\bf W}_{i}^{\rm I}\right\|^2+{\sigma _{{k}}^2}},\forall k,m.\label{snr}
\end{equation}
Thus, the 
consumed transmitted power of the $k$th ISCC device in the first time slot is $E_{t,k}^{\rm I}={\rm tr}\left({\overline{\bf W}}_{k}^{\rm I}\right)t_{k}^{\rm I}$.

In the second time slot $t_{k}^{\rm II}$, the ISCC devices forward the remote computation data via the communication stream to the AP, while the sensing stream is used for spatial registration of the cooperative sensing area in preparation for the next frame transmission. 

The offloading rate of the $k$th ISCC device at the AP can be expressed as
\begin{equation}
				\setlength{\abovedisplayskip}{3pt}
	\setlength{\belowdisplayskip}{3pt}
	r_k^{\rm II}=\log_2\det\left({\bf I}_G+{\bf H}_k{\overline{\bf W}}_{c,k}^{\rm II}{\bf H}_k^H\left({\bf R}_k^{\rm II}\right)^{-1}\right),\forall k,
\end{equation}
where ${\bf R}_k^{\rm II}\!=\!\sigma_{n_k}^2{\bf I}_N\!+\sum\nolimits_{i \neq k}^K{\bf H}_k{\overline{\bf W}}_{c,i}^{\rm II}{\bf H}_k^H+\sum\nolimits_{i = 1}^K{\bf H}_k{\overline{\bf W}}_{s,i}^{\rm II}{\bf H}_k^H$. ${\bf x}^{\rm II}_k$ is defined similarly to ${\bf x}^{\rm I}_k$. Correspondingly, the offloading delay of the $k$th ISCC device should satisfy the following
\begin{equation}
	\setlength{\abovedisplayskip}{3pt}
	\setlength{\belowdisplayskip}{3pt}
	t_{k}^{\rm II}\ge{d_{k}^p}/{r_k^{\rm II}},\forall k.
\end{equation}
The sensing constraint on the second time slot will be described below. Thus, the 
consumed transmitted power of the $k$th ISCC device in the second time slot is $E_{t,k}^{\rm II}={\rm tr}\left({\overline{\bf W}}_{k}^{\rm II}\right)t_{k}^{\rm II}$.
\subsection{Computation Model}
\subsubsection{Data Acquisition Model} The quantity of data that the network needs to execute stems from the sensing data obtained in the first time slot, which in this paper we define the obtained sensing data as the following form
\begin{equation}
					\setlength{\abovedisplayskip}{3pt}
	\setlength{\belowdisplayskip}{3pt}
	D_{k,m}=f_s q_bSINR^{echo}_{k,m}t_{k}^{\rm I},\forall k,m,
\end{equation}
where $f_s$ and $q_b$ denote the sampling frequency and quantization bit, respectively. The acquisition of sensing data is also related to the radar beam switching speed and quantization coefficient\cite{2012INTRODUCTION}, which we describe in this paper using a simplified model. Meanwhile, the SINR of the echo signal is introduced, which is a consideration of the influence of the interference and clutter in the echo on the effective data. On the selection of network performance metrics, we consider the $D_{k}=\sum_{m=1}^MD_{k,m}$ as one of the metrics. This represents not only the ability of sensing to acquire valid data, but also the amount of data that the network is able to compute, and on the other hand, the capacity that the communication can support.

\subsubsection{ISCC Devices Local Computation Model} As mentioned in the previous subsection, we divide the data $D_{k}$ into two parts for computation, which are the ISCC device local computation $d_{k}^l$ and the AP remote computation $d_{k}^r$. For the local computation model, it can be expressed as follows
\begin{equation}
					\setlength{\abovedisplayskip}{3pt}
	\setlength{\belowdisplayskip}{3pt}
	f_{l,i}\le f_{l,max},\forall i\in\left\{1,2,\cdots,c_ld_{k}^l\right\},\label{fm}
\end{equation}
where $f_{l,i}$, $f_{l,max}$ and $c_l$ denote the CPU frequency for the $i$th cycle, the maximum CPU computing frequency and the number of CPU cycles for computing each one task input-bit at the ISCC device, respectively.
In order for the local computation to be completed within $T$, the following computational latency constraints therefore need to be satisfied
\begin{equation}
					\setlength{\abovedisplayskip}{3pt}
	\setlength{\belowdisplayskip}{3pt}
	\sum_{i=1}^{c_ld_{k}^l}\frac{1}{f_{l,i}}\le T-t_{k}^{\rm I},\forall k.\label{ft}
\end{equation}
For minimizing the energy consumption, the power consumed is minimized when all frequencies are equal and the inequality takes an equal sign in Eq. (\ref{ft}), and combining this with Eq. (\ref{fm}), there are \cite{8234686}
\begin{equation}
					\setlength{\abovedisplayskip}{3pt}
	\setlength{\belowdisplayskip}{3pt}
	\frac{c_ld_{k}^l}{T-t_{k}^{\rm I}}\le f_{l,max},\forall k.
\end{equation}
Therefore, under the assumption that the CPU low-voltage condition generally holds, the local computation energy consumption of the $k$th ISCC device can be expressed as \cite{burd1996processor}
\begin{equation}
					\setlength{\abovedisplayskip}{3pt}
	\setlength{\belowdisplayskip}{3pt}
	E_{l,k}=\sum_{i=1}^{c_ld_{k}^l}\alpha_lf_{l,i}^2=\frac{\alpha_l\left(c_ld_{k}^l\right)^3}{\left(T-t_{k}^{\rm I}\right)^2},\forall k,
\end{equation}
where $\alpha_l$ denotes the effective capacitance coefficient that depends on the chip architecture at the ISCC device.
\subsubsection{MEC Remote Computation Model} In time slot $t_{k}^{\rm III}$, the MEC server executes the remote computation task. Similarly, the time required for the MEC to execute the input bit $d_{k}^r$ is
\begin{equation}
	t_{k}^{\rm III} = {c_rd_{k}^r}/{f_{r,i}},\forall i\in\left\{1,2,\cdots,c_rd_{k}^r\right\},
\end{equation}
where ${f_{r,i}}$ and $c_r$ denote the CPU frequency for the $i$th cycle and the CPU frequency for the $i$th cycle for computing each one task input-bit at the MEC, respectively. The MEC server runs at maximum computational frequency and needs to satisfy the following constraints
\begin{equation}
					\setlength{\abovedisplayskip}{3pt}
	\setlength{\belowdisplayskip}{3pt}
	\sum_{k=1}^{K}\frac{c_rd_{k}^r}{t_{k}^{\rm III}}\le f_{r,max},\forall k,
\end{equation}
where ${f_{r,max}}$ denotes the maximum CPU computing frequency at the MEC. Therefore, the remote compution energy consumption at the MEC with respect to the $k$th ISCC device can be expressed as 
\begin{equation}
					\setlength{\abovedisplayskip}{3pt}
	\setlength{\belowdisplayskip}{3pt}
	E_{r,k}=\sum_{i=1}^{c_rd_{k}^r}\alpha_rf_{r,i}^2=\frac{\alpha_r\left(c_rd_{k}^r\right)^3}{\left(t_{k}^{\rm III}\right)^2},\forall k,
\end{equation}
where $\alpha_r$ denotes the effective capacitance coefficient that depends on the chip architecture at the MEC server.
\subsection{Multi-source Information Fusion Mechanisms}
Since the proposed ISCC architecture requires the cooperation of multiple devices to achieve high-accuracy sensing, multi-source information fusion at multiple levels needs to be considered. Fusion is subdivided into signal-level fusion and data-level fusion. Signal-level fusion coherently fuses the echo signals from different BSs to improve the SNR of the echo signals, and it is required to match the sensing areas of different BSs with higher accuracy\cite{9800700}. However, unaligned sensing regions of different ISCC devices can lead to signal spatial mismatch and weaken the fusion effect. Spatial registration can be achieved by beamwidth adjustment, which not only improves the energy gain of signal fusion, but also reduces the impact of synchronization error on parameter estimation, so this paper focuses on this level.
\begin{figure*}[ht] 
	{\centerline{\includegraphics[width=18cm]{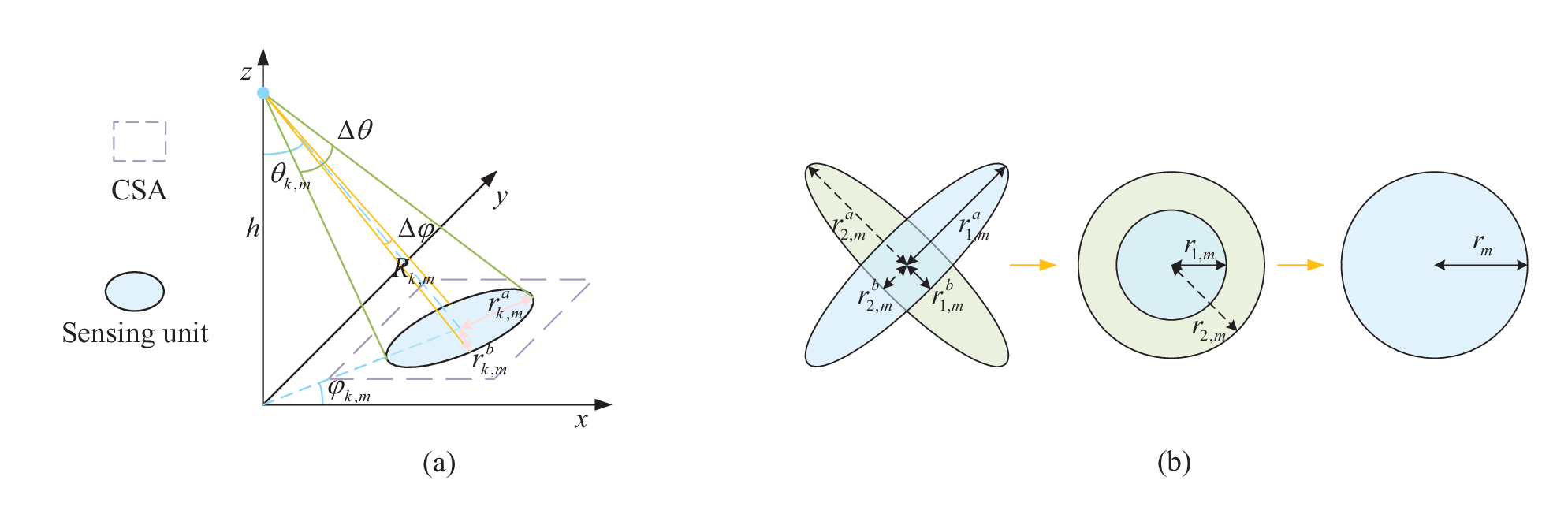}}}
	\caption{Schematic diagram of beam coverage and beamwidth adjustment.}\label{BA}
\end{figure*}

The cooperative sensing area (CSA) can be approximated as a plane parallel to the ground, which can be expressed in polar coordinate system as
\begin{equation}
					\setlength{\abovedisplayskip}{3pt}
	\setlength{\belowdisplayskip}{3pt}
	r\cos\theta=h-R_{k,m}\cos\theta_{k,m},\label{csa}
\end{equation}
where $h$ denotes the height of the ISCC device, $R_{k,m}$ denotes the distance between the target and the ISCC device, and $\left(\theta_{k,m},\varphi_{k,m}\right)$ denotes the angle of the beam center.

For the case of small and distant targets, the beam can be approximated as an oblique cylinder with an elliptical base, which can be represented in the polar coordinate system as
\begin{equation}	
	\frac{\left(r\sin{\theta_{rot}}\cos\varphi\right)^2}{\tan^2\left({\Delta\theta_{k,m}}/{2}\right)}+\frac{\left(r\sin{\theta_{rot}}\sin\varphi\right)^2}{\tan^2\left({\Delta\varphi_{k,m}}/{2}\right)}=R_{k,m}^2,\label{fb}
\end{equation}
where $\left(\Delta\theta_{k,m},\Delta\varphi_{k,m}\right)$ denotes the beamwidth and $\theta_{rot}=\theta-\theta_{k,m}-\pi/2$ denotes the projection of the angle between the beam center and the target in the horizontal plane. 

Combining (\ref{csa}) and (\ref{fb}) to solve, we get that the sensing area is an ellipse whose long and short axes can be expressed as
\begin{equation}
					\setlength{\abovedisplayskip}{3pt}
	\setlength{\belowdisplayskip}{3pt}
	r^a_{k,m}={R_{k,m}\tan\left({\Delta\theta_{k,m}}/{2}\right)}/{\cos\theta_{k,m}},
\end{equation}
and
\begin{equation}
					\setlength{\abovedisplayskip}{3pt}
	\setlength{\belowdisplayskip}{3pt}
	r^b_{k,m}={R_{k,m}\tan\left({\Delta\varphi_{k,m}}/{2}\right)}.
\end{equation}
In order to make the cooperative sensing units consistent, we first approximate the sensing units as circles, i.e., $r^a_{k,m} = r^b_{k,m}$. Then there are
\begin{equation}
					\setlength{\abovedisplayskip}{3pt}
	\setlength{\belowdisplayskip}{3pt}
	\Delta\theta_{k,m}=2{\rm arctan}\left(\cos\theta_{k,m}\tan\left({\Delta\varphi_{k,m}}/{2}\right)\right).\label{Dthe}
\end{equation}
Next to make all sensing units of equal size, we rank the target distances $R_{k,m}$ acquired in the first stage and use the median $\left(R_0,\theta_0,\varphi_0,\Delta\theta_0,\Delta\varphi_0\right)$ as a reference. Then adjusting the beamwidth according to Eqs. (\ref{Dthe}) and (\ref{Dfai}), the size of the sensing unit will be the same.
\begin{equation}
	R_{k,m}\tan\left({\Delta\varphi_{k,m}}/{2}\right)=R_{0}\tan\left({\Delta\varphi_0}/{2}\right).\label{Dfai}
\end{equation}
In the following, we describe how to adjust the beamwidth to achieve spatial registration, and a schematic of the above process is shown in Fig. \ref{BA}. 

In the collaborative scenario, adjusting the beam width is a key technology to realize the spatial registration of the sensing area. By adjusting the beam width, the range and shape of the sensing area of a single BS can be changed, so that the sensing areas of different BSs can spatially overlap, and ultimately enhance the overall performance of the network. In the following we introduce a beamwidth adjustable algorithm.

Firstly, we define angle vectors ${\bm\theta}=\left[\theta_0,\cdots,\theta_0+\left(L-1\right)\Delta_1\right]\in\mathbb{C}^{L\times1}$ and ${\bm\varphi}=\left[\varphi_0,\cdots,\varphi_0+\left(L-1\right)\Delta_2\right]\in\mathbb{C}^{L\times1}$, with elements spaced $\Delta_1$ and $\Delta_2$, respectively. The desired directional amplitude response of the $k$th ISCC device can be expressed as
\begin{equation}
	\begin{split}
		{ r}_{k}^{ad}\left({\theta_l},{\varphi_l}\right)=\sum\limits_{e=1}^{E}\frac{2}{1+e^{\left(\left(\frac{\theta_l-\theta_{e}}{{\Delta \theta}/2}\right)^2+\left(\frac{\varphi_l-\varphi_{e}}{{\Delta \varphi}/2}\right)^2\right)\times{\ln\left(2\sqrt{2}-1\right)}}},
	\end{split}
\end{equation}
where $A=\left\{\left(\theta_e,\varphi_e\right)|e\in\left\{1,\cdots,M,AP\right\}\right\}$ denotes the set of all desired signal directions, and we take the direction of AP into account. $\left(\Delta\theta,\Delta\varphi\right)$ denotes the expected beam pitch and azimuth width. To simplify the processing, we set the beam widths of the different targets to be the same. The constant $ln(2\sqrt{2}-1)$ is used to limit the -3 dB half-power beamwidth. For subsequent processing, we reorganize $r_{ad}$ in angle-indexed order into the matrix ${\bf R}_{k}^{ad}\in\mathbb{C}^{L\times L}$.

Secondly, we define the steering matrix of the $k$th ISCC device as
\begin{equation}
	\begin{split}
		{\bf A}_k=&\left[{\bf a}_1\left(\theta_{k,1},\varphi_{k,1}\right),\cdots,{\bf a}_l\left(\theta_{k,l},\varphi_{k,l}\right),\right.\\
		&\left.{\bf a}_L\left(\theta_{k,L},\varphi_{k,L}\right)\right]\in\mathbb{C}^{N_t\times L},\forall k,
	\end{split}
\end{equation}
where ${\bf a}_l\left(\cdot\right)$ denotes the steering vector.

Then, the directional response of the antenna after beamforming can be expressed as
\begin{equation}
	{\bf r}_k=({\bf W}_{k}^{\rm II})^H{\bf A}_k, \forall k.
\end{equation}

Finally, in order to ensure the spatial registration performance in the second time slot, the following constraint need to be satisfied
\begin{equation}
\sum_{k=1}^K\left\|{\bf r}_k^{H}{\bf r}_k-{\bf R}_{k}^{ad}\right\|^2\le \gamma_{th},
\end{equation}
where $\gamma_{th}$ denotes the threshold of spatial registration error.
\subsection{Problem Formulation}
The management of power consumption in future networks is a complex problem, in this paper the network involves transmit energy and computational energy, so we consider minimizing the total energy consumption of the network by jointly optimizing the beamforming matrixs ${\overline{\bf W}}_{k}^{\rm S}={\overline{\bf W}}_{c,k}^{\rm S}+{\overline{\bf W}}_{s,k}^{\rm S}$, time allocation $\bf T$, sensing data ${\bf D}$, and sensing beam scheduling variables ${\bm \rho}$. The optimization problem can be written as
\begin{subequations}
	\begin{align}
		&\left( {{\textrm{P0}}} \right){\textrm{:~}}{\mathop {\textrm{min}~}\limits_{{\overline{\bf W}}_{c,k}^{\rm S},{\overline{\bf W}}_{s,k}^{\rm S},{\bf T},{\bf D},{\bm \rho}}}~\sum_{k=1}^K\left(E_{t,k}^{\rm I}+E_{t,k}^{\rm II}+E_{l,k}+E_{r,k}\right), \notag\\
		&~~{\textrm{s}}{\textrm{.t}}{\rm{.}}~~~\left[{\overline{\bf W}}_{k}^{\rm S}\right]_{nn}\le P_t, {\rm S}\in\left\{{\rm I,II}\right\},\forall k,n,\\
		&~~~~~~~~~{\overline{\bf W}}_{c,k}^{\rm S}\succeq 0,{\rm S}\in\left\{{\rm I,II}\right\},\forall k,\\
		&~~~~~~~~~{\overline{\bf W}}_{s,k}^{\rm S}\succeq 0,{\rm S}\in\left\{{\rm I,II}\right\},\forall k,\\
		&~~~~~~~~~{\rm rank}({\overline{\bf W}}_{c,k}^{\rm S})\le G,{\rm S}\in\left\{{\rm I,II}\right\},\forall k,\\
		&~~~~~~~~~\sum_{k=1}^K\left\|{\bf A}_k^H{\overline{\bf W}}_{k}^{\rm II}{\bf A}_k-{\bf R}_{k}^{ad}\right\|^2\le \gamma_{th},\\
		&~~~~~~~~~r_k^{\rm I}\ge R_{th},\forall k,\label{rth}\\
		&~~~~~~~~~t_{k}^{\rm S}\ge 0, S\in\left\{{\rm I,II,III}\right\}, \forall k,\\
        &~~~~~~~~~t_{k}^{\rm I}+t_{k}^{\rm II}+t_{k}^{\rm III}\le T, \forall k,\\
		&~~~~~~~~~\frac{c_ld_{k}^l}{T-t_{k}^{\rm I}}\le f_{l,max},\forall k,\\
		&~~~~~~~~~\sum_{k=1}^{K}\frac{c_rd_{k}^r}{t_{k}^{\rm III}}\le f_{r,max},\\
		&~~~~~~~~~d_{k}^r\le t_{k}^{\rm II}r_k^{\rm II},\forall k,\\
		&~~~~~~~~~d_{k}^l+d_{k}^r= D_{k},\forall k,\label{Dk}\\
		&~~~~~~~~~1\le\sum_{m=1}^{M}\rho_{k,m}\le M,\forall k,\label{ro1}\\
		&~~~~~~~~~1\le \sum_{k=1}^K\rho_{k,m}\le K,\forall m,\label{ro2}\\
		&~~~~~~~~~\rho_{k,m}\in \left\{0,1\right\},\forall k,m,
 	\end{align}
\end{subequations}
where Eq. (\ref{rth}) denotes the achievable rate constraint. Eqs. (\ref{ro1})-(\ref{ro2}) indicates that an ISCC device can sense multiple targets. However, the optimization variables are highly coupled, resulting in the objective function and constraints being nonconvex with respect to the optimization variables, and the constraints (\ref{ro1})-(\ref{ro2}) involve binary variables. As a result, problem (P0) is a mixed-integer nonconvex optimization problem, which is quite challenging to obtain a globally optimal solution. Therefore, we need to design an efficient algorithm to obtain a high-quality suboptimal solution via the BCD algorithm in the next section.

\section{Joint Optimization Algorithm For the ISCC Network}
In this section, since problem (P0) is a non-convex optimization problem, we divide it into three subproblems to solve, based on the BCD algorithm. 
\subsection{Block I: Sensing Beam Scheduling Strategy}
In this subsection, given the beamforming matrixs ${\overline{\bf W}}_{k}^{\rm S}$, the time allocation $\bf T$ and the sensing data $\bf D$, the sensing beam scheduling variables ${\bm \rho}$ is optimized, which is a feasible check problem. However, the presence of the binary variable $\rho_{k,m}$ causes constraints (\ref{Dk})-(\ref{ro2}) to be non-convex, and to address this, we relax the sensing beam scheduling variable $\rho_{k,m}\in\left\{0,1\right\}$ to $\widetilde\rho_{k,m}\in\left[0,1\right]$. To deal with constraint (\ref{Dk}), we introduce the auxiliary variable $p_{k,m}$. After variable handling, the optimization problem (P0) can be expressed as the optimization problem (P1) as follows
\begin{subequations}
				\setlength{\abovedisplayskip}{3pt}
	\setlength{\belowdisplayskip}{3pt}
	\begin{align}
		&\left( {{\textrm{P1}}} \right){\textrm{:~}}{\mathop {\textrm{find}~}}~{\bm \rho}, \notag\\
		&~~{\textrm{s}}{\textrm{.t}}{\rm{.}}~~~~d_{k}^l+d_{k}^r\ge f_sq_bt_k^{\rm I}\sum\limits_{m=1}^Mp_{k,m},\forall k,\label{dd}\\
		&~~~~~~~~~p_{k,m}\ge\frac{\widetilde\rho_{k,m}\sum\limits_{i=1}^K\left\|{\bf z}_{k,m,i}{\bf W}_{i}^{\rm I}\right\|^2}{\sum\limits_{j\neq m}^M\widetilde\rho_{k,j}\sum\limits_{i=1}^K\left\|{\bf z}_{k,j,i}{\bf W}_{i}^{\rm I}\right\|^2+\sigma^2_{n_k}},\forall k,m,\label{fra}\\
		&~~~~~~~~~1\le \sum_{m=1}^{M}\widetilde\rho_{k,m}\le M,\forall k,\label{rm}\\
		&~~~~~~~~~1\le \sum_{k=1}^K\widetilde\rho_{k,m}\le K,\forall m,\\
		&~~~~~~~~~\widetilde\rho_{k,m}\in\left[0,1\right],\forall k,m.\label{rin}
	\end{align}
\end{subequations}
We first transform the fractional constraint (\ref{fra}) into the following form
\begin{align}
				\setlength{\abovedisplayskip}{3pt}
	\setlength{\belowdisplayskip}{3pt}
		&\widetilde\rho_{k,m}\sum\limits_{i=1}^K{\bf z}_{k,m,i}{\overline{\bf W}}_{i}^{\rm I}{\bf z}_{k,m,i}^H-p_{k,m}\sum\limits_{j\neq m}^M\widetilde\rho_{k,j}\sum\limits_{i=1}^K{\bf z}_{k,j,i}{\overline{\bf W}}_{i}^{\rm I}{\bf z}_{k,j,i}^H\notag\\
		&-p_{k,m}\sigma^2_{n_k}\le0,\forall k,m.\label{pr}
\end{align}
Then problem (P1) can be decoupled as follows
\begin{subequations}
				\setlength{\abovedisplayskip}{3pt}
	\setlength{\belowdisplayskip}{3pt}
	\begin{align}
		&\left( {{\textrm{P1.1}}} \right){\textrm{:~}}{\mathop {\textrm{find}~}}~{\bm \rho}, \notag\\
		&~~~~~~~~~~~{\textrm{s}}{\textrm{.t}}{\rm{.}}~~{\textrm{(\ref{rm})-(\ref{rin}),(\ref{pr})}}. 
	\end{align}
\end{subequations}
Given $p_{k,m}$, problem (P1.1) is a feasible checking problem that can be easily solved using the CVX toolbox\cite{cvxtool}. After solving problem (P1.1) to obtain $\bm \rho$, then there are
\begin{subequations}
				\setlength{\abovedisplayskip}{3pt}
	\setlength{\belowdisplayskip}{3pt}
	\begin{align}
		&\left( {{\textrm{P1.2}}} \right){\textrm{:~}}{\mathop {\textrm{find}~}}~{\bm p}, \notag\\
		&~~~~~~~~~~~{\textrm{s}}{\textrm{.t}}{\rm{.}}~~{\textrm{(\ref{dd})-(\ref{fra})}}.
	\end{align}
\end{subequations}
Similarly, problem (P1.2) is a feasible checking problem that can be easily solved.

{\it Remark: To facilitate the solution of problem {\rm (P1)}, we relax the sensing beam scheduling variable $\rho_{k,m}$ into a continuous variable $\widetilde\rho_{k,m}$, which will lead to a suboptimal solution with high accuracy. After solving problem {\rm (P1)}, we recover $\rho_{k,m}$ using an empirical threshhold scheme, assigning 1 to variables larger than the threshold and 0 to the rest.}

\subsection{Block II: Communication and Sensing Beamforming Design}
In this subsection, given the time allocation $\bf T$, the sensing data $\bf D$, the sensing beam scheduling variables ${\bm \rho}$ and the auxiliary variable $\bf p$, the beamforming matrixs ${\overline{\bf W}}_{k}^{\rm S}$ are optimized, then the problem (P0) can be transformed into the problem (P2), which can be expressed as
\begin{subequations}
				\setlength{\abovedisplayskip}{3pt}
	\setlength{\belowdisplayskip}{3pt}
	\begin{align}
		&\left( {{\textrm{P2}}} \right){\textrm{:~}}{\mathop {\textrm{min}~}\limits_{{\overline{\bf W}}_{c,k}^{\rm S},{\overline{\bf W}}_{s,k}^{\rm S}}}~f({\overline{\bf W}}_{c,k}^{\rm S},{\overline{\bf W}}_{s,k}^{\rm S}), \notag\\
		&~~{\textrm{s}}{\textrm{.t}}{\rm{.}}~~~\left[{\overline{\bf W}}_{k}^{\rm S}\right]_{nn}\le P_t, {\rm S}\in\left\{{\rm I,II}\right\},\forall k,n,\label{pp}\\
		&~~~~~~~~~{\overline{\bf W}}_{c,k}^{\rm S}\succeq 0,{\rm S}\in\left\{{\rm I,II}\right\},\forall k,\\
		&~~~~~~~~~{\overline{\bf W}}_{s,k}^{\rm S}\succeq 0,{\rm S}\in\left\{{\rm I,II}\right\},\forall k,\\
		&~~~~~~~~~\sum_{k=1}^K\left\|{\bf A}_k^H{\overline{\bf W}}_k^{\rm II}{\bf A}_k-{\bf R}_{k}^{ad}\right\|^2\le \gamma_{th},\label{nr}\\
		&~~~~~~~~~{\rm rank}({\overline{\bf W}}_{c,k}^{\rm S})\le G,{\rm S}\in\left\{{\rm I,II}\right\},\forall k,\label{rankG}\\
		&~~~~~~~~~r_k^{\rm I}\ge R_{th},\forall k,\label{Rk}\\
		&~~~~~~~~~d_{k}^r\le t_{k}^{\rm II}r_k^{\rm II},\forall k,\label{rk}\\
		&~~~~~~~~~d_{k}^l+d_{k}^r\ge \widetilde{D}_{k},\forall k,\label{snd}
	\end{align}
\end{subequations}
where $f({\overline{\bf W}}_{c,k}^{\rm S},{\overline{\bf W}}_{s,k}^{\rm S})=\sum_{k=1}^K\left({\rm tr}\left({\overline{\bf W}}_{k}^{\rm I}\right)t_k^{\rm I}+{\rm tr}\left({\overline{\bf W}}_{k}^{\rm II}\right)t_k^{\rm II}\right.\\$
$\left.+E_{l,k}+E_{r,k}\right)$.
Since $r_k^{\rm S}$ is not a convex constraint in constraints (\ref{Rk}) and (\ref{rk}) with respect to the variable ${\bf W}_{k}^{\rm S}$, we first transform it as follows
\begin{equation}
				\setlength{\abovedisplayskip}{3pt}
	\setlength{\belowdisplayskip}{3pt}
\begin{split}
	r_k^{\rm S}= \log_2{\rm det}\left(\sigma_{n_k}^2{\bf I}_N+\sum_{i=1}^K{\bf H}_k{\overline{\bf W}}_{i}^{\rm S}{\bf H}_k^H\right)\\
	-\log_2{\rm det}\left({\bf R}_k^{\rm S}\right),\forall k.
\end{split}
\end{equation}
Then by utilizing successive convex approximations (SCA), the upper bound of $\log_2{\rm det}\left({\bf R}_k\right)$ can be expressed as (\ref{ubb}),
 \begin{figure*}[ht] 
	\centering
	\begin{equation}
		\begin{split}
			\log_2{\rm det}\left({\bf R}_k^{\rm S}\right)^{ub}\triangleq \log_2{\rm det}\left({\bf R}_k^{{\rm S}\left(r\right)}\right)&+{\rm tr}\left(\left({\bf R}_k^{{\rm S}(r)}\right)^{-1}\left(\sum_{i\neq k}^K{\bf H}_k\left({\overline{\bf W}}_{c,i}^{\rm S}-{\overline{\bf W}}_{c,i}^{{\rm S}(r)}\right){\bf H}_k^H\right.\right.\\
			&\left.\left.+\sum_{i=1}^K{\bf H}_k\left({\overline{\bf W}}_{s,i}^{\rm S}-{\overline{\bf W}}_{s,i}^{{\rm S}(r)}\right){\bf H}_k^H\right)\right),\forall k.
		\end{split}\label{ubb}
	\end{equation}
	\hrulefill
\end{figure*}
where $r$ denotes the number of iterations. By replacing $\log_2{\rm det}\left({\bf R}_k^{\rm S}\right)$ with $\log_2{\rm det}\left({\bf R}_k^{\rm S}\right)^{ub}$, constraint (\ref{Rk}) and (\ref{rk}) are both transformed into convex constraints as follow
\begin{equation}
				\setlength{\abovedisplayskip}{3pt}
	\setlength{\belowdisplayskip}{3pt}
	\begin{split}
	&\log_2{\rm det}\left(\sigma_{n_k}^2{\bf I}_N+\sum_{k=1}^K{\bf H}_k{\overline{\bf W}}_{k}^{\rm I}{\bf H}_k^H\right)\\
&~~~~~~~~~~~~~~~~-\log_2{\rm det}\left({\bf R}_k^{\rm I}\right)^{ub}\ge R_{th},\forall k,
	\end{split}\label{s1}
\end{equation}
and 
\begin{equation}
				\setlength{\abovedisplayskip}{3pt}
	\setlength{\belowdisplayskip}{3pt}
	\begin{split}
	&\log_2{\rm det}\left(\sigma_{n_k}^2{\bf I}_N+\sum_{k=1}^K{\bf H}_k{\overline{\bf W}}_{k}^{\rm II}{\bf H}_k^H\right)\\
&~~~~~~~~~~~~~~~~-\log_2{\rm det}\left({\bf R}_k^{\rm II}\right)^{ub}\ge d_k^r/t_k^{\rm II},\forall k.
	\end{split}\label{s2}
\end{equation}
For constraint (\ref{snd}), we have transformed it into the constraints (\ref{dd}) and (\ref{pr}). To handle the constraint (\ref{rankG}), we first introduce {\bf Proposition 1}.
\begin{prop}
The $(g+1)$th largest eigenvalue $\lambda_{n-g}$ of matrix ${\bf U} \in \mathbb{S}_+^n$ is no greater than $e$ if and only if $e{\bf I}_{n-g} - {\bf V}^T{\bf U}{\bf V} \succeq {\bf 0}$, where ${\bf I}_{n-g}$ is the identity matrix with a dimension of $n-g$, ${\bf V} \in \mathbb{R}^{n \times (n-g)}$ are the eigenvectors corresponding to the $n-g$ smallest eigenvalues of ${\bf U}.$ When $e=0$ and ${\bf U}$ is a positive semidefinite matrix, ${\rm rank}({\bf U}) \leq g$ holds if and only if $e{\bf I}_{n- g}- {\bf V}^T{\bf U}{\bf V}\succeq {\bf 0}.$ 
\end{prop}
{\it Proof:} Assume the eigenvalues of ${\bf U}$ is sorted in descending orders in the form of $[\lambda_n,\lambda_{n-1},\ldots,\lambda_1].$ Since the Rayleigh quotient of an eigenvector is its associated eigenvalue, then $e{\bf I}_{n-g}-{\bf V}^T{\bf U}{\bf V}$ is a diagonal matrix with diagonal elements set as $[e-\lambda_{n-g},e-\lambda_{n-g-1},\ldots,e-\lambda_1].$ Therefore $e\geq\lambda_{n-g}$ if and only if $e{\bf I}_{n-g}-{\bf V}^T{\bf U}{\bf V}\succeq {\bf 0}.$ $\hfill\blacksquare$

Since the eigenvector ${\bf V}_k$ about ${\overline{\bf W}}_k^{\rm S}$ is unknown, we introduce an iterative scheme to approximate the rank gradually. Then the problem (P2) can be expressed as
\begin{subequations}
				\setlength{\abovedisplayskip}{3pt}
	\setlength{\belowdisplayskip}{3pt}
	\begin{align}
		&\left( {{\textrm{P2.1}}} \right){\textrm{:~}}{\mathop {\textrm{min}~}\limits_{{\overline{\bf W}}_{c,k}^{\rm S},{\overline{\bf W}}_{s,k}^{\rm S},e_k}}~f({\overline{\bf W}}_{c,k}^{\rm S},{\overline{\bf W}}_{s,k}^{\rm S})+\sum_{\rm S}\sum_{k=1}^{K}c_0c^re_k^{\rm S}, \notag\\
		&~~~{\textrm{s}}{\textrm{.t}}{\rm{.}}~~~{\textrm{(\ref{pr}),}}{\textrm{(\ref{pp})-(\ref{nr}),}}{\textrm{(\ref{s1})-(\ref{s2}),}}\\
		&~~~~~~~~~e_k^{\rm S}{\bf I}_{N_t- g}- \left({\bf V}_k^T\right)^{(r)}{\overline{\bf W}}_{c,k}^{\rm S}{\bf V}_k^{(r)}\succeq {\bf 0},\forall k,\\
		&~~~~~~~~~e_k^{\rm S}\le e_k^{{\rm S}(r)},\forall k,
	\end{align}
\end{subequations}
where $c_0\textgreater 1$ denotes the initial regularization factor, $c^r(c\textgreater 1)$ is increasing with the increment of iteration $r$, and ${\bf V}_k^{(r)}$ are the orthonormal eigenvectors corresponding to the $N_t-g$ smallest eigenvalues of ${\overline{\bf W}}_{c,k}^{\rm S}$ at $r$th iteration. The constraint $e_k^{\rm S}\le e_k^{{\rm S}(r)}$ guarantees that $e_k^{\rm S}$ is monotonically decreasing in the
iterative algorithm. Until $e_k^{\rm S}$ decreases to $0$, according to {\bf Proposition 1}, then ${\rm rank}({\bf W}_{c,k}^{\rm S})\le G$ holds. This scheme is referred to as iterative rank minimization \cite{sun2018rank,10464353}. At this point, all the non-convex terms of problem (P2.1) have been transformed into convex ones and it is a typical SDP problem that can be handed over to the CVX toolbox to be solved.

\subsection{Block III: Joint Optimization of Sensing Data and Time Allocation}
In this subsection, given the beamforming matrixs ${\bf W}_{k}^{\rm S}$ and the sensing beam scheduling variables ${\bm \rho}$ and the auxiliary variable $\bf p$, the time allocation $\bf T$ and the sensing data $\bf D$ are jointly optimized, then the problem (P0) can be transformed into the problem (P3), which can be expressed as
\begin{subequations}
				\setlength{\abovedisplayskip}{3pt}
	\setlength{\belowdisplayskip}{3pt}
	\begin{align}
		&\left( {{\textrm{P3}}} \right){\textrm{:~}}{\mathop {\textrm{min}~}\limits_{{\bf T},{\bf D}}}~\sum_{k=1}^K\left({\rm tr}\left({\overline{\bf W}}_{k}^{\rm I}\right)t_k^{\rm I}+{\rm tr}\left({\overline{\bf W}}_{k}^{\rm II}\right)t_k^{\rm II}\notag\right.\\
		&~~~~~~~~~~~~~~~~~~~~~\left.+\frac{\alpha_l\left(c_ld_{k}^l\right)^3}{\left(T-t_{k}^{\rm I}\right)^2}+\frac{\alpha_r\left(c_rd_{k}^r\right)^3}{\left(t_{k}^{\rm III}\right)^2}\right),\notag\\
		&~~~~~~~~~{\textrm{s}}{\textrm{.t}}{\rm{.}}~~t_{k}^{\rm S}\ge 0, S\in\left\{{\rm I,II,III}\right\}, \forall k,\label{tt1}\\
		&~~~~~~~~~~~~~~t_{k}^{\rm I}+t_{k}^{\rm II}+t_{k}^{\rm III}\le T, \forall k,\label{tt2}\\
		&~~~~~~~~~~~~~~\frac{c_ld_{k}^l}{T-t_{k}^{\rm I}}\le f_{l,max},\forall k,\label{fl}\\
		&~~~~~~~~~~~~~~\sum_{k=1}^{K}\frac{c_rd_{k}^r}{t_{k}^{\rm III}}\le f_{r,max},\label{fr}\\
		&~~~~~~~~~~~~~~d_{k}^r\le t_{k}^{\rm II}r_k^{\rm II},\forall k,\label{tt3}\\
		&~~~~~~~~~~~~~~d_{k}^l+d_{k}^r\ge \widetilde{D}_{k},\forall k.\label{tt4}
	\end{align}
\end{subequations}

We first deal with the objective function $E_{l,k}$ and $E_{r,k}$, and introduce the following $\textbf{Lemma 1. }$
\begin{lemma}
	$f_1\left(x,y\right)= \frac {x ^{3}}{\left ( c- y\right ) ^{2}}$ and $f_2\left(x,y\right)= \frac {x ^{3}}{  y ^{2}}$ are convex functions with respect to $x>0$ and $y>0$, where $c>y$ is a positive constant.
\end{lemma}
${Proof}{:}$ The Hessian matrix of $f_1\left(x,y\right)$ is
\begin{equation}
				\setlength{\abovedisplayskip}{3pt}
	\setlength{\belowdisplayskip}{3pt}
\left.\left[\begin{array}{cc}\frac{6x}{\left(c-y\right)^2}&\frac{6x^2}{\left(c-y\right)^3}\\\frac{6x^2}{\left(c-y\right)^3}&\frac{6x^3}{\left(c-y\right)^4}\end{array}\right.\right].
\end{equation}
Its eigenvalues are 0 and $\frac{6x}{{\left(c-y\right)}^2}\left(1+\frac{x^2}{{\left(c-y\right)}^2}\right)$ , so the matrix is a semipositive definite matrix, then $f_1$ is jointly convex with respect to $x>0$ and $y>0$.

The Hessian matrix of $f_2\left(x,y\right)$ is
\begin{equation}
				\setlength{\abovedisplayskip}{3pt}
	\setlength{\belowdisplayskip}{3pt}
	\left.\left[\begin{array}{cc}\frac{6x}{y^2}&-\frac{6x^2}{y^3}\\-\frac{6x^2}{y^3}&\frac{6x^3}{y^4}\end{array}\right.\right].
\end{equation}
Its eigenvalues are 0 and $\frac{6x}{{y}^2}\left(1+\frac{x^2}{{y}^2}\right)$ , so the matrix is a semipositive definite matrix, then $f_2$ is jointly convex with respect to $x>0$ and $y>0$. This completes the proof of
$\textbf{Lemma 1. }$ $\hfill\blacksquare$

By utilizing $\textbf{Lemma 1}$, we can obtain that the eigenvalues of $E_{l,k}$ are 0 and $\frac{6c_l^3\alpha_ld_k^c}{\left(T-t_k^\mathrm{I}\right)^2}\left(1+\frac{(d_k^l)^2}{\left(T-t_k^\mathrm{I}\right)^2}\right)$, and hence $E_{l,k}$ is jointly convex with respect to $d_k^l>0$ and $t_k^\mathrm{I}>0$. The eigenvalues of $E_{r,k}$ are 0 and $\frac{6c_r^3\alpha_rd_k^r}{\left(t_k^\mathrm{III}\right)^2}\left(1+\frac{(d_k^r)^2}{\left(t_k^\mathrm{III}\right)^2}\right)$, and hence $E_{r,k}$ is jointly convex with respect to $d_k^r>0$ and $t_k^\mathrm{III}>0$.

Next we deal with constraints (\ref{fl}) and (\ref{fr}), the constraints (\ref{fl}) can be transformed into 
\begin{equation}
				\setlength{\abovedisplayskip}{3pt}
	\setlength{\belowdisplayskip}{3pt}
	c_ld_k^l+t_k^{\rm I}f_{l,max}-Tf_{l,max}\le 0,\forall k.\label{tT}
\end{equation}
Transform the item to the left of constraint (\ref{fr}) into the following form utilizing SCA.
\begin{equation}
				\setlength{\abovedisplayskip}{3pt}
	\setlength{\belowdisplayskip}{3pt}
	\begin{split}
		\left(\frac{c_rd_k^{r}}{t_k^{{\rm III}}}\right)^{ub}&\triangleq\frac{c_rd_k^{r{(r)}}}{t_k^{{\rm III}{(r)}}}+\frac{c_r}{t_k^{{\rm III}{(r)}}}\left(d_k^r-d_k^{r{(r)}}\right)\\
		&-\frac{c_rd_k^{r{(r)}}}{\left(t_k^{{\rm III}{(r)}}\right)^2}\left(t_k^{\rm III}-t_k^{{\rm III}{(r)}}\right),\forall k.
	\end{split}
\end{equation}
Thus, the constraint (\ref{fr}) can be rewritten as 
\begin{equation}
				\setlength{\abovedisplayskip}{3pt}
	\setlength{\belowdisplayskip}{3pt}
	\sum_{k=1}^K\left(\frac{c_rd_k^{r}}{t_k^{{\rm III}}}\right)^{ub}\le f_{r,max}.\label{tm}
\end{equation}
Then the problem (P3) can be transformed into the problem (P3.1), which can be expressed as
\begin{subequations}
				\setlength{\abovedisplayskip}{3pt}
	\setlength{\belowdisplayskip}{3pt}
	\begin{align}
	&\left( {{\textrm{P3.1}}} \right){\textrm{:~}}{\mathop {\textrm{min}~}\limits_{{\bf T},{\bf D}}}~\sum_{k=1}^K\left({\rm tr}\left({\overline{\bf W}}_{k}^{\rm I}\right)t_k^{\rm I}+{\rm tr}\left({\overline{\bf W}}_{k}^{\rm II}\right)t_k^{\rm II}\notag\right.\\
	&~~~~~~~~~~~~~~~~~~~~\left.+\frac{\alpha_l\left(c_ld_{k}^l\right)^3}{\left(T-t_{k}^{\rm I}\right)^2}+\frac{\alpha_r\left(c_rd_{k}^r\right)^3}{\left(t_{k}^{\rm III}\right)^2}\right),\notag\\
		&~~~~~~~~~~~{\textrm{s}}{\textrm{.t}}{\rm{.}}~~~{\textrm{(\ref{tt1}),(\ref{tt2}),(\ref{tt3}),(\ref{tt4}),(\ref{tT}),(\ref{tm}).}}
	\end{align}
\end{subequations}
After dealing with all the constraints and objective function, problem (P3.1) is a standard convex optimization problem that can be solved using the CVX toolbox.
\subsection{The Joint Optimization Algorithm for Spatial Registration Based on Adjustable Beamwidth in ISCC Networks}
The original problem is a non-convex optimization problem due to the high coupling of multiple optimization variables. In this paper, we decompose it into three subproblems to solve. Specifically, in block I, given the beamforming matrix ${\overline{\bf W}}_k^{{\rm S}}$, the time allocation $\bf T$, and the sensing data $\bf D$, the sensing beam scheduling variable $\bm \rho$ is optimized by linear programming. In block II, given the time allocation $\bf T$, the sensing data $\bf D$, and the sensing beam scheduling variable $\bm \rho$, optimize the beamforming matrix ${\overline{\bf W}}_k^{{\rm S}}$, via SDP. In block III, given the beamforming matrix ${\overline{\bf W}}_k^{{\rm S}}$ and the sensing beam scheduling variable $\bm \rho$, jointly optimize the sensing data $\bf D$ as well as the time allocation $\bf T$, via SCA. The three sub-problems are solved iteratively until convergence and summarized in the following {\bf Algorithm 1}\footnote{The initial feasible solution for the proposed algorithm is obtained by selecting any solution that satisfies the constraints. For complex constraints, a projection mapping approach is employed, followed by relaxation and subsequent adjustment of the solution. When solving subproblems, the fixed variables in other blocks are already constrained. Optimization is performed directly within the feasible region of the current block, requiring only that the current block's variables satisfy their corresponding constraints. The resulting solution naturally forms a globally feasible solution together with the fixed variables in other blocks.}.
\begin{algorithm}[htbp]
	\caption{The Joint Optimization Algorithm for Spatial Registration Based on Adjustable Beamwidth in ISCC Networks}
	\label{alg1}
	\begin{algorithmic}[1]
		\STATE {\bf{Initialization}}: ${\overline{\bf W}}_k^{{\rm S}(0)}$, ${\bf T}^{(0)}$, ${\bf D}^{(0)}$, ${\bm \rho}^{(0)}$, ${\bf p}^{(0)}$, convergence threshold $\varepsilon$ and iteration index $r = 0$.
		\REPEAT
		\STATE Obtain sensing beam scheduling $\bm \rho$ by solving the problem (P1.1) and (P1.2).
		
		\STATE Obtain beamforming ${\overline{\bf W}}_k^{{\rm S}}$ by solving the problem (P2.1).
		
		\STATE Obtain sensing data ${\bf D}$ and time allocation $\bf T$ by solving the problem (P3.1).
		\STATE $r \leftarrow r + 1$.
		\UNTIL The fractional decrease of the objective value is
		below a threshold $\varepsilon$.
		\STATE {\bf{return}} The sensing beam scheduling, beamforming, sensing data and time slot allocation design scheme.
	\end{algorithmic}  
\end{algorithm}
\subsection{Computational Complexity and Convergence Analysis}
\subsubsection{Computational Complexity Analysis}
In each iteration, the problems (P1.1) and (P1.2) are solved with computational complexity ${\cal O}\left(KM\right)$ and ${\cal O}\left(KM\right)$, respectively. The problem (P2.1) is a SDP problem solved with complexity ${\cal O}\left(\left(KN_t\right)^{3.5}\right)$. The problem (P3.1) is sovled  with computational complexity ${\cal O}\left(K^{3.5}\right)$. Therefore, the overall computational complexity of {\bf Algorithm 1} is ${\cal O}\left(r_{max}\left(\left(KN_t\right)^{3.5}+K^{3.5}+2KM\right)\right)$, where $r_{max}=\log\left(1/\varepsilon\right)$ is the number of iterations required for the algorithm to converge and $\varepsilon$ is the precision of stopping the iteration.
\subsubsection{Convergence Analysis}
The convergence of the proposed joint optimized beamforming, sensing beam scheduling, sensing data and time slot allocation is elaborated as follows. For given ${\overline{\bf W}}_k^{{\rm S}(r)},{\bf T}^{(r)},{\bf D}^{(r)}$, we obtain the sensing beam scheduling ${\bm \rho}^{(r)}$ and the auxiliary variable ${\bf p}^{(r)}$, and the objective function ${\cal F}\left({\overline{\bf W}}_k^{{\rm S}(r)},{\bf T}^{(r)},{\bf D}^{(r)},{\bm \rho}^{(r)},{\bf p}^{(r)}\right)$ in the step 3 of the {\bf Algorithm 1} has 
\begin{equation}
			\setlength{\abovedisplayskip}{3pt}
	\setlength{\belowdisplayskip}{3pt}
	\begin{split}
	{\cal F}\left({\overline{\bf W}}_k^{{\rm S}(r)},{\bf T}^{(r)},{\bf D}^{(r)},{\bm \rho}^{(r)},{\bf p}^{(r)}\right)\ge \\
	{\cal F}\left({\overline{\bf W}}_k^{{\rm S}(r)},{\bf T}^{(r)},{\bf D}^{(r)},{\bm \rho}^{(r+1)},{\bf p}^{(r+1)}\right).
	\end{split}
\end{equation}
For given ${\bf T}^{(r)},{\bf D}^{(r)},{\bm \rho}^{(r+1)},{\bf p}^{(r+1)}$, we obtain the beamforming matrixs ${\overline{\bf W}}_k^{{\rm S}(r)}$, and the objective function ${\cal F}\left({\overline{\bf W}}_k^{{\rm S}(r)},{\bf T}^{(r)},{\bf D}^{(r)},{\bm \rho}^{(r+1)},{\bf p}^{(r+1)}\right)$ in the step 4 of the {\bf Algorithm 1} has 
\begin{equation}
				\setlength{\abovedisplayskip}{3pt}
	\setlength{\belowdisplayskip}{3pt}
	\begin{split}
		{\cal F}\left({\overline{\bf W}}_k^{{\rm S}(r)},{\bf T}^{(r)},{\bf D}^{(r)},{\bm \rho}^{(r+1)},{\bf p}^{(r+1)}\right)\ge \\
		{\cal F}\left({\overline{\bf W}}_k^{{\rm S}(r+1)},{\bf T}^{(r)},{\bf D}^{(r)},{\bm \rho}^{(r+1)},{\bf p}^{(r+1)}\right).
	\end{split}
\end{equation}
For given ${\overline{\bf W}}_k^{{\rm S}(r+1)},{\bm \rho}^{(r+1)},{\bf p}^{(r+1)}$, we obtain the sensing data ${\bf D}^{(r)}$ and time slot allocation ${\bf T}^{(r)}$, and the objective function ${\cal F}\left({\overline{\bf W}}_k^{{\rm S}(r+1)},{\bf T}^{(r)},{\bf D}^{(r)},{\bm \rho}^{(r+1)},{\bf p}^{(r+1)}\right)$ in the step 4 of the {\bf Algorithm 1} has 
\begin{equation}
				\setlength{\abovedisplayskip}{3pt}
	\setlength{\belowdisplayskip}{3pt}
	\begin{split}
		{\cal F}\left({\overline{\bf W}}_k^{{\rm S}(r+1)},{\bf T}^{(r)},{\bf D}^{(r)},{\bm \rho}^{(r+1)},{\bf p}^{(r+1)}\right)\ge \\
		{\cal F}\left({\overline{\bf W}}_k^{{\rm S}(r+1)},{\bf T}^{(r+1)},{\bf D}^{(r+1)},{\bm \rho}^{(r+1)},{\bf p}^{(r+1)}\right).
	\end{split}
\end{equation}
Based on the above derivation, we have
\begin{equation}
				\setlength{\abovedisplayskip}{3pt}
	\setlength{\belowdisplayskip}{3pt}
	\begin{split}
		{\cal F}\left({\overline{\bf W}}_k^{{\rm S}(r)},{\bf T}^{(r)},{\bf D}^{(r)},{\bm \rho}^{(r)},{\bf p}^{(r)}\right)\ge \\
		{\cal F}\left({\overline{\bf W}}_k^{{\rm S}(r+1)},{\bf T}^{(r+1)},{\bf D}^{(r+1)},{\bm \rho}^{(r+1)},{\bf p}^{(r+1)}\right).
	\end{split}
\end{equation}
The above equation shows that the objective function of {\bf Algorithm 1} is non-increasing at each iteration and the objective function must be a lower bound for a finite value, so the convergence of {\bf Algorithm 1} can be guaranteed.
\section{Numerical Results}
In this section, numerical simulations of the proposed algorithm are performed to demonstrate its effectiveness. The channel implementation employs a multiple-sample averaging approximation. A three-dimensional polar coordinate system is used, the AP is located at (0{\rm{m}}, -50{\rm{m}}, 10{\rm{m}}), $K=3$ ISCC devices are distributed at coordinates (-50r\rm{m}, 0\rm{m}, 10\rm{m}), (0\rm{m}, 0\rm{m}, 10\rm{m}), and (50\rm{m}, 0\rm{m}, 10\rm{m}), respectively, and $M=5$ DUEs are distributed with coordinates of (-75\rm{m}, 50\rm{m}, 0\rm{m}), (-25\rm{m}, 50\rm{m}, 0\rm{m}), (0\rm{m}, 50\rm{m}, 0\rm{m}), (25\rm{m}, 50\rm{m}, 0\rm{m}), and (75\rm{m}, 50\rm{m}, 0\rm{m}), respectively. The remaining parameters are given in Table I.
\begin{table}[!htbp]
	\caption{Simulation Parameters}
	\begin{center}
		\begin{tabular}{|l|c|}  
			\hline
			\bf{Parameters} & \bf{Value}\\ 
			\hline
			Carrier frequency ($f$) & 30 GHz \\
			\hline
			System bandwidth ($W$) & 20 MHz \\
			\hline
			Noise power (${\sigma_{n_k} ^2}$, ${\sigma_{n_m} ^2}$,${\sigma_{n_u} ^2}$) & -90 dBm \\
			\hline
			Convergence precision ($\varepsilon$) & $10^{ - 3}$\\
			\hline
			ISCC equivalent capacitance factor ($\alpha_l$) & $0.3\times10^{ - 27} A^2s^4/F$\\
			\hline
			ISCC CPU cycle/bit ($c_l$) & $10^{3}$\\
			\hline
			Maximum frequency of the ISCC CPU ($f_{lmax}$) & 3 GHz\\
			\hline
			MEC equivalent capacitance factor ($\alpha_r$) & $1\times10^{ - 27} A^2s^4/F$\\
			\hline
			MEC CPU cycle/bit ($c_r$) & $10^{3}$ cycle/bit\\
			\hline
			Maximum frequency of the MEC CPU ($f_{rmax}$) & 5 GHz\\
			\hline
		\end{tabular}
	\end{center}
\end{table}

In this paper, we compare the performance of the proposed algorithm and other benchmarks as follows: (1) {\it{ISCC Local Computation}} (ILC): in this case, the ISCC device locally computes all the acquired sensing data in time $T-t_k^{\rm I}$. (2) {\it{MEC Remote Computation}} (MRC): in this case, the ISCC device offloads all acquired sensing data to the MEC for computation, with the ISCC device acting as a relay to assist in the offloading. (3) {\it{Traditional Cooperative Transceiver}} (TCT): this scheme uses traditional multi-antenna transceivers for ISCC and the network operates in a cooperative manner with a power constraint that can be expressed as ${\rm tr}\left({\overline{\bf W}}_k^{\rm S}\right)\le N_tP_{t}$. (4) {\it{Non-cooperative}} (NC): the scenario network operates in a stand-alone manner and does not consider sensing beam scheduling. (5) {\it{Imperfect registration}} (IR): the CSA of this scheme is not fully aligned.

\begin{figure}[!htpb]
	\centerline{\includegraphics[width=5.5cm]{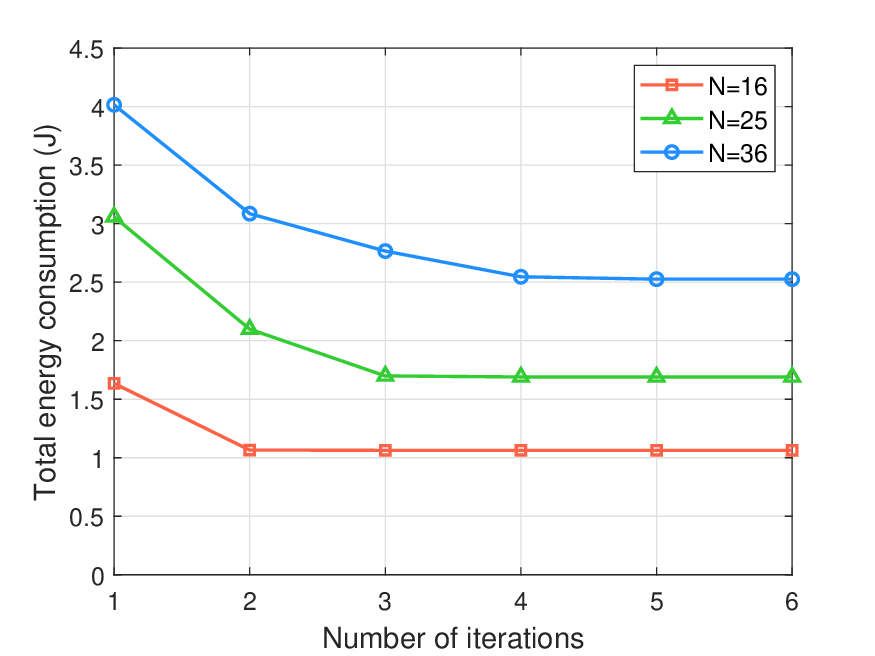}}
	\caption{Convergence process: Total energy consumption versus number of iterations under different TRIS elements ($P_t = 1 {\rm W}, T=1{\rm s}$).}\label{conv}
\end{figure}
First, the convergence of the proposed algorithm is demonstrated in Fig. \ref{conv}, where it can be seen that the number of iterations needed for convergence increases from 2 to 4 as the number of TRIS elements increases, and the total energy consumption of the network increases.
\begin{figure}[!htbp]
	\centering
	\begin{minipage}[t]{1\linewidth} 
		\centering
		\begin{tabular}{@{\extracolsep{\fill}}c@{}c@{}@{\extracolsep{\fill}}}
			\includegraphics[width=0.52\linewidth]{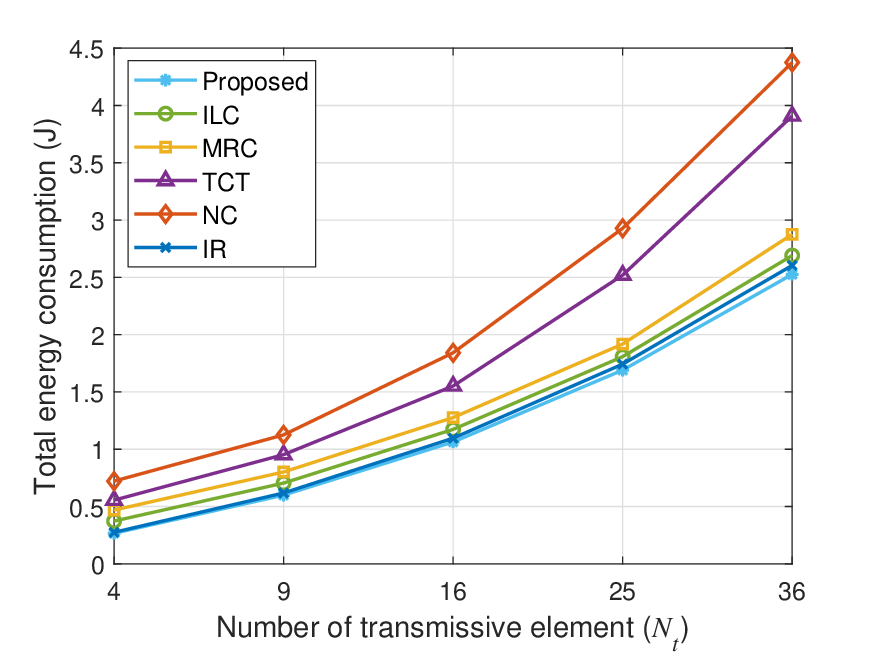}&
			\includegraphics[width=0.52\linewidth]{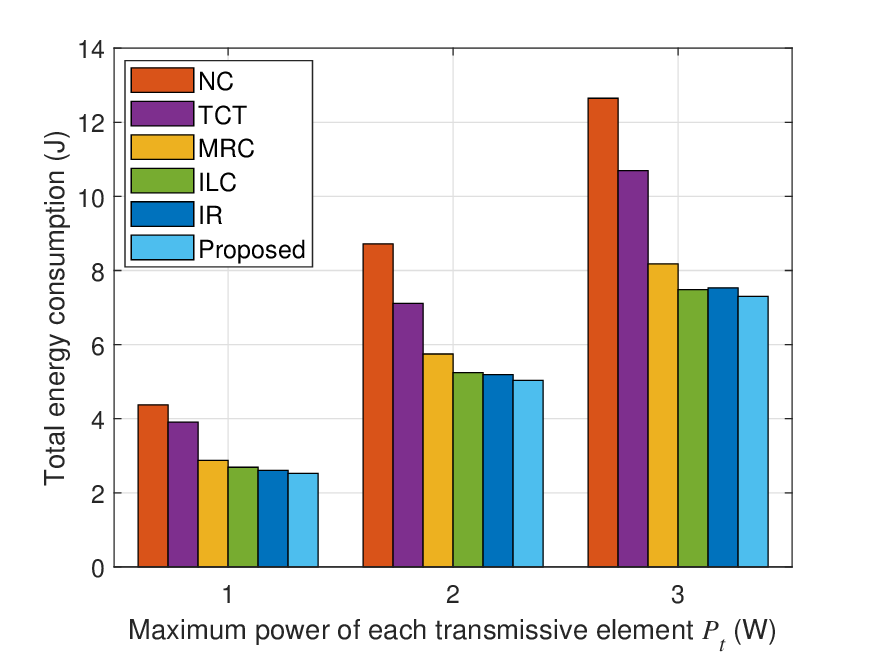}\\
			(a)  & (b) \\
		\end{tabular}
	\end{minipage}
	\caption{Total energy consumption varies with the number of TRIS element and the maximum power of each transmissive element ($N_t=36, T=1{\rm s}, P_{md}=0.97, P_t=1{\rm W}$): (a) TEC v.s. $N_t$, (b) TEC v.s. $P_t$.}\label{EPN}
\end{figure}

\begin{figure}[!htbp]
	\centering
	\begin{minipage}[t]{1\linewidth} 
		\centering
		\begin{tabular}{@{\extracolsep{\fill}}c@{}c@{}@{\extracolsep{\fill}}}
			\includegraphics[width=0.52\linewidth]{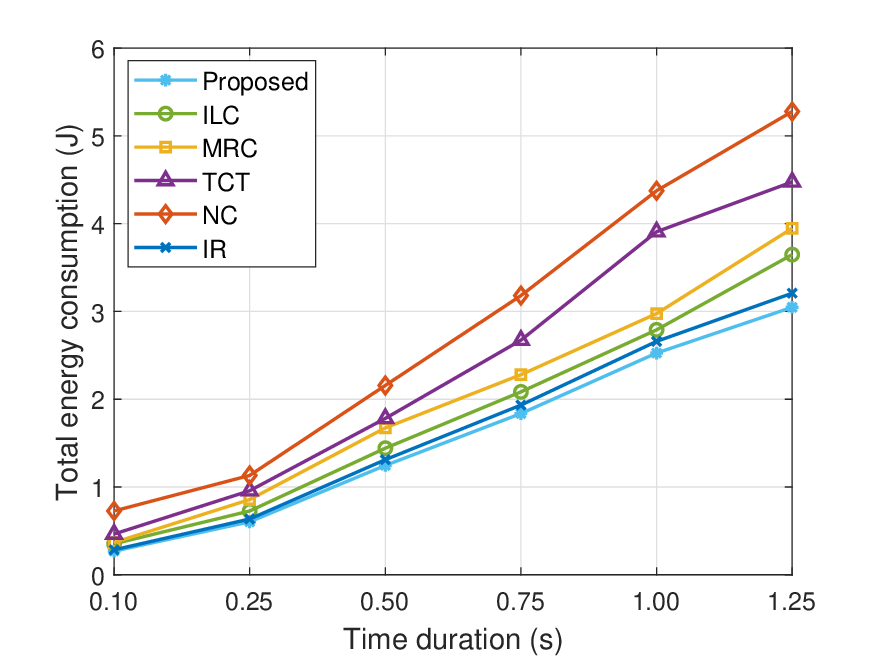}&
			\includegraphics[width=0.52\linewidth]{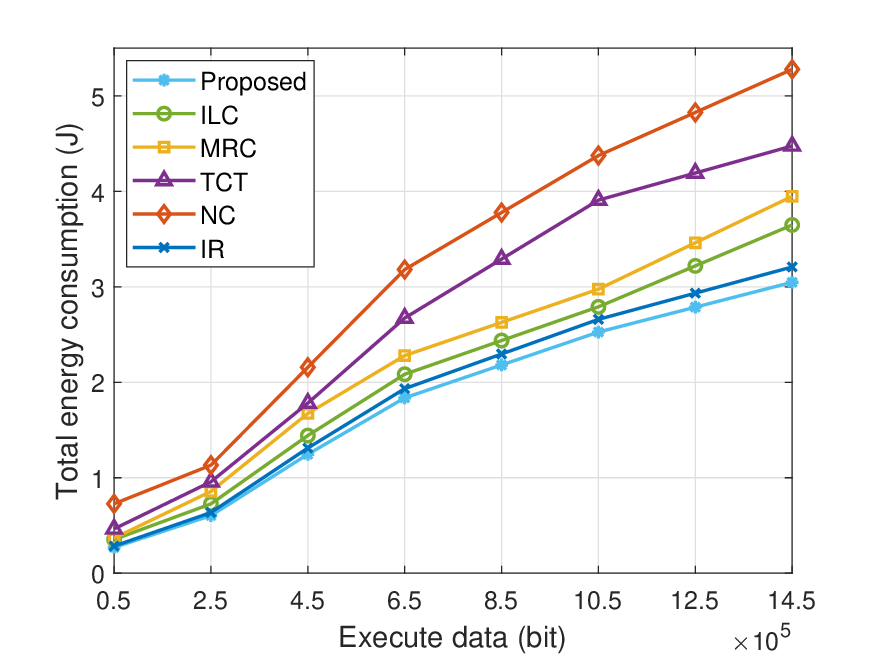}\\
			(a) TEC v.s. $T$ & (b) TEC v.s. ED\\
		\end{tabular}
	\end{minipage}
	\caption{Total energy consumption varies with the time duration and the execute data ($N_t=36, P_t=1{\rm W}, P_{md}=0.97$).}\label{EDT}
\end{figure}

Next, we consider the total energy consumption (TEC) with different parameters, in Fig. \ref{EPN}, the total energy consumption increases with increasing number of TRIS elements and with the maximum power of the TRIS elements. In addition, according to Fig. \ref{EDT}, the total energy consumption tends to increase as the time duration and the amount of data to be executed increases, which is due to the fact that the network has more data to be executed and therefore consumes more energy. Taken as a whole, the energy consumption of the proposed scheme is reduced by 35.36\% as compared to the traditional transceivers, which proves the superiority of TRIS transceivers in terms of energy saving. The 42.26\% energy consumption reduction compared to the non-cooperative scheme proves that the cooperative scheme brings cooperative gains in terms of energy consumption, which is due to the fact that interference cancellation and sensing beam scheduling are not considered in the non-cooperative mode, resulting in the need for greater transmit energy to achieve the desired SINR constraint. The energy consumption is reduced by 12.15\% and 6.11\% compared with the remote computing and local computing solutions, respectively. This is due to the more powerful computing capability at the MEC, and the designed two- layer computing is able to effectively reduce the network energy consumption through the rational allocation of tasks and the optimization of resource synergy. Compared with the imperfect registration scheme, the energy consumption rises by only 3\% with 97\% matching, which demonstrates the robustness of the proposed scheme.
\begin{figure}[!htbp]
	\centering
	\begin{minipage}[t]{1\linewidth} 
		\centering
		\begin{tabular}{@{\extracolsep{\fill}}c@{}c@{}@{\extracolsep{\fill}}}
			\includegraphics[width=0.52\linewidth]{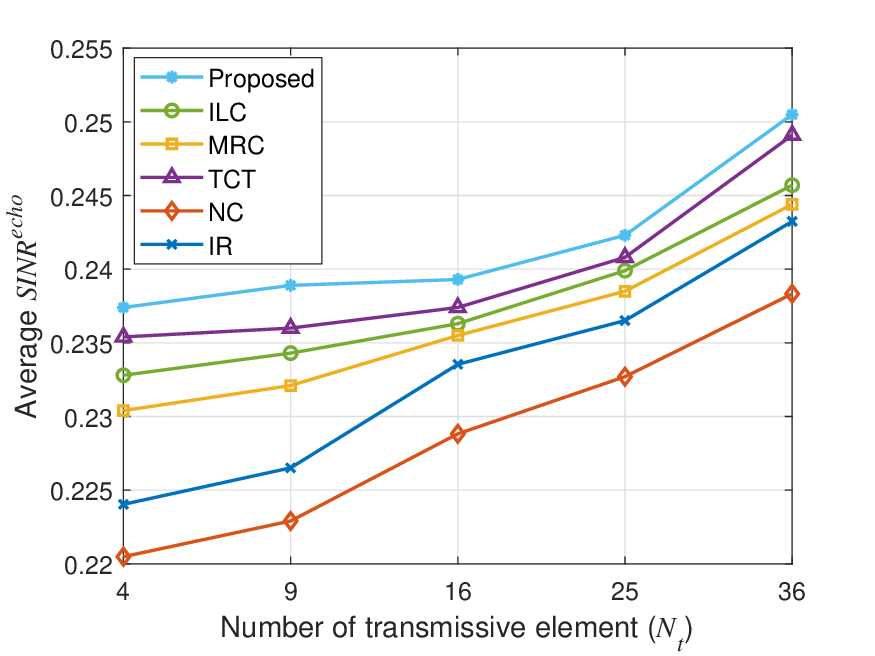}&
			\includegraphics[width=0.52\linewidth]{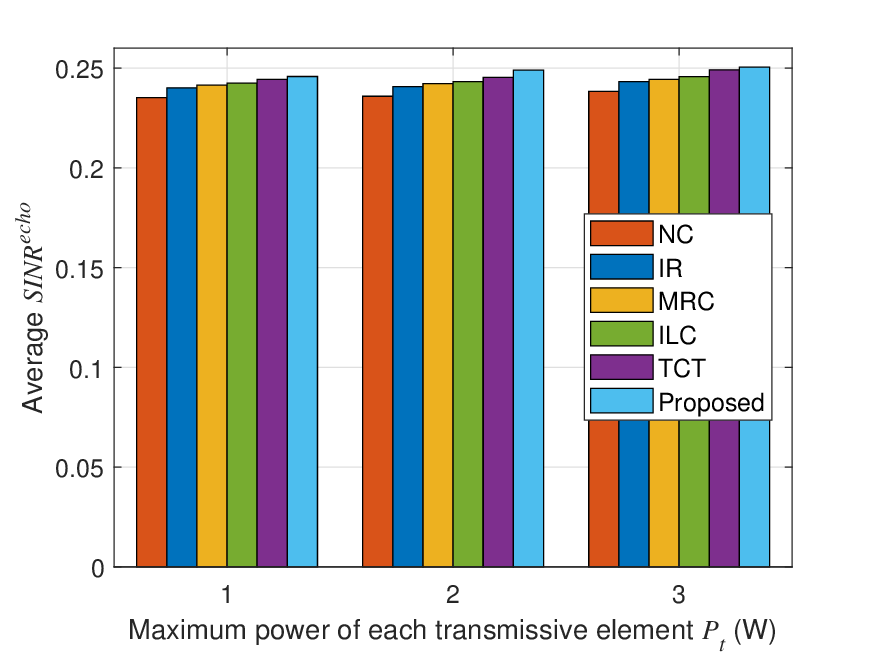}\\
			(a)  & (b) \\
		\end{tabular}
	\end{minipage}
	\caption{Average echo SINR varies with the number of TRIS element and the maximum power of each transmissive element ($N_t=36, T=1{\rm s}, P_{md}=0.97, P_t=3{\rm W}$): (a) ${SINR}^{echo}$ v.s. $N_t$, (b) ${SINR}^{echo}$ v.s. $P_t$.}\label{APN}
\end{figure}

Then, the variation of the average echo SINR with the TRIS parameters is demonstrated in Fig. \ref{APN}. As the number of TRIS elements increases, the interference of the signal is better suppressed and the useful signal is enhanced, resulting in a significant increase in the ${SINR}^{echo}$. Meanwhile, by increasing the maximum power of each transmissive element, the echo energy is enhanced and thus the ${SINR}^{echo}$ is also enhanced. The sensing performance of the proposed scheme is enhanced compared to traditional transceivers due to the fact that TRIS is able to modulate the beam in a simpler and flexible way. The 5\% enhancement compared to the NC is due to the higher interference in the non-cooperative condition. The ILC performance is higher than MRC due to the fact that ILC does not require data offloading and more power is allocated to the sensing beam. It can also be seen that the IR has a greater impact on the SINR, when the cooperative sensing region is not perfectly registered, and the effective echo signal is affected by scattering and clutter, resulting in a decrease in the SINR. The reason for the low SINR improvement is that TRIS is equipped with only a single receive antenna with low receive gain.

\begin{figure}[!htpb]
	\centerline{\includegraphics[width=5.5cm]{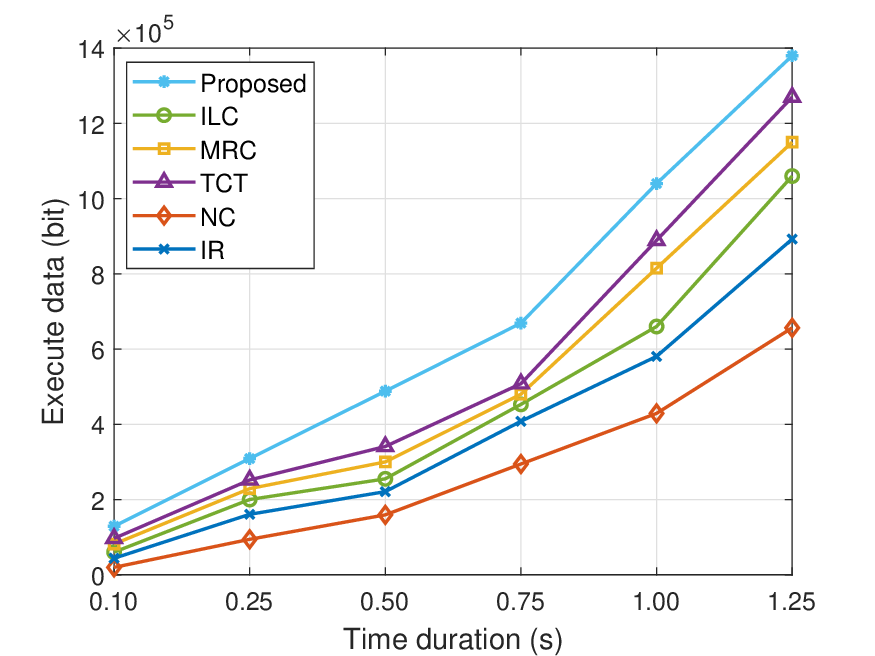}}
	\caption{Execute data varies with the time duration ($N_t=36, P_t=1{\rm W}, P_{md}=0.97$).}\label{DT}
\end{figure}
 
Further, it can be seen in Fig. \ref{DT} that the amount of execute data tends to increase as the time duration $T$ increases, this is due to the fact that an increase in $T$ will cause the first time slot $t_k^{\rm I}$ to increase and the ISCC device will acquire more sensing data. Comparing with the baseline, the network-supported execute data enhancement ranges from 8.66\% to 110\%, indicating that the proposed architecture can well support large-scale data computation and transmission.
\begin{figure}[!htbp]
	\centering
	\begin{minipage}[t]{1\linewidth} 
		\centering
		\begin{tabular}{@{\extracolsep{\fill}}c@{}c@{}@{\extracolsep{\fill}}}
			\includegraphics[width=0.52\linewidth]{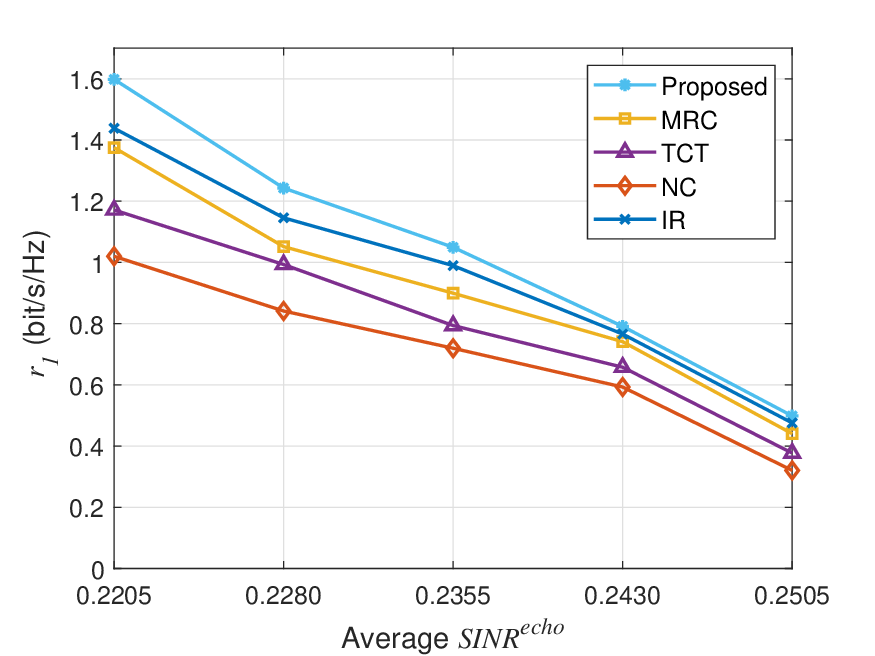}&
			\includegraphics[width=0.52\linewidth]{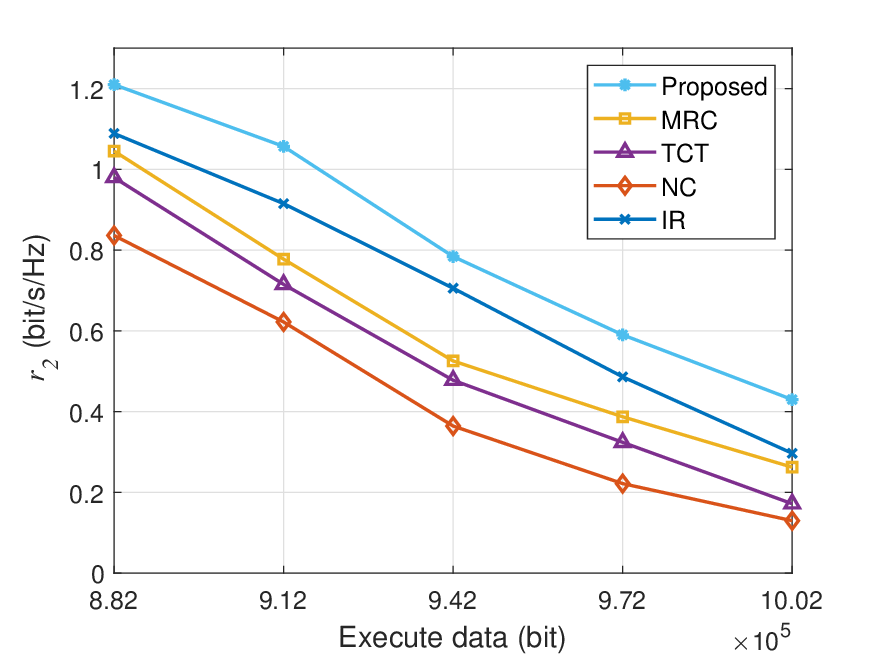}\\
			(a) $r_1$ v.s. ${SINR}^{echo}$ & (b) $r_2$ v.s. ED\\
		\end{tabular}
	\end{minipage}
	\caption{Sum-rate varies with the average echo SINR and the execute data ($P_t=1{\rm W}, T=1{\rm s}, N_t=36, P_{md}=0.97$).}\label{RSD}
\end{figure}

This is followed by a discussion of the tradeoffs between communication and sensing as well as computation. As shown in Fig. \ref{RSD}, the communication rate tends to decrease with both average echo $SINR$ and execute data $D$. This is due to the fact that as $SINR$ and $D$ increase, more energy resources are allocated to sensing and computation, which results in less energy being allocated to communication, and thus a decrease in the offloading rate. In addition, the offloading rate of the proposed scheme is 23.47\% higher than that of TCT, which indicates that TRIS has higher spectral efficiency and is more suitable for multi-stream communication.
\begin{figure}[!htbp]
	\centering
	\begin{minipage}[t]{1\linewidth} 
		\centering
		\begin{tabular}{@{\extracolsep{\fill}}c@{}c@{}@{\extracolsep{\fill}}}
			\includegraphics[width=0.52\linewidth]{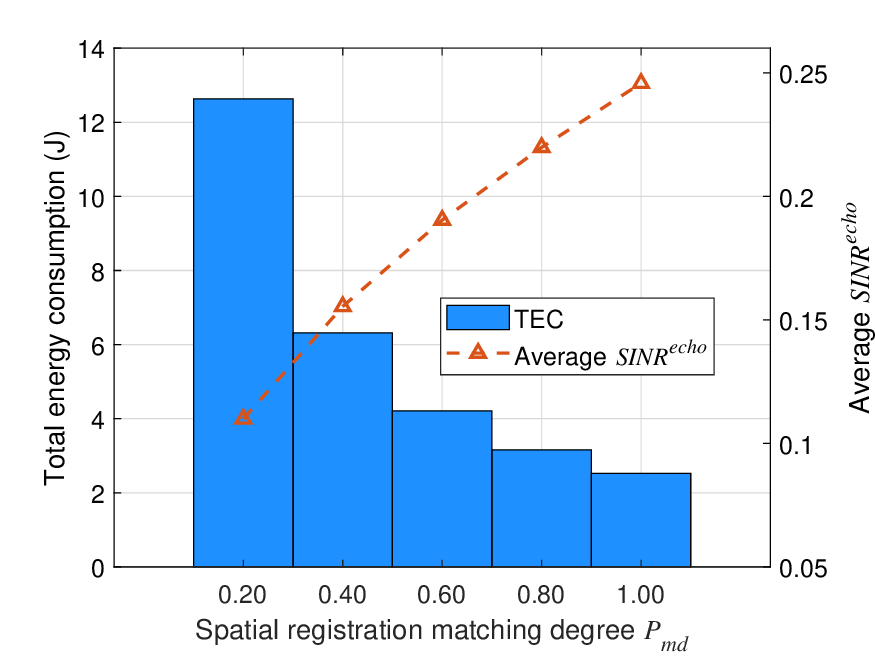}&
			\includegraphics[width=0.52\linewidth]{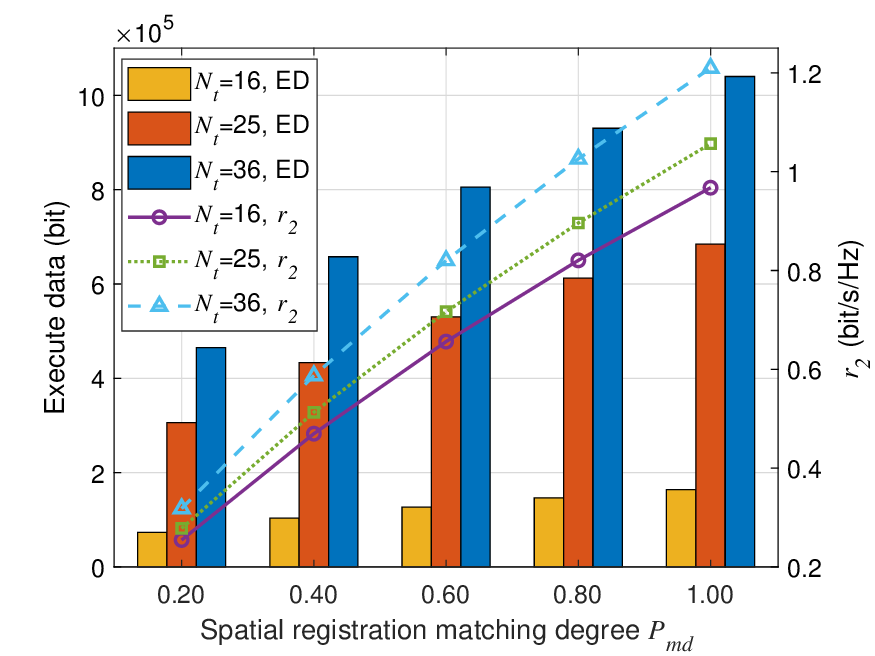}\\
			(a) & (b) \\
		\end{tabular}
	\end{minipage}
	\caption{Performance variation with different spatial registration degree of match ($P_t=1{\rm W}, T=1{\rm s}, N_t=36$).}\label{tw}
\end{figure}

\begin{figure}[!htbp]
	\centering
	\begin{minipage}[t]{1\linewidth} 
		\centering
		\begin{tabular}{@{\extracolsep{\fill}}c@{}c@{}@{\extracolsep{\fill}}}
			\includegraphics[width=0.53\linewidth]{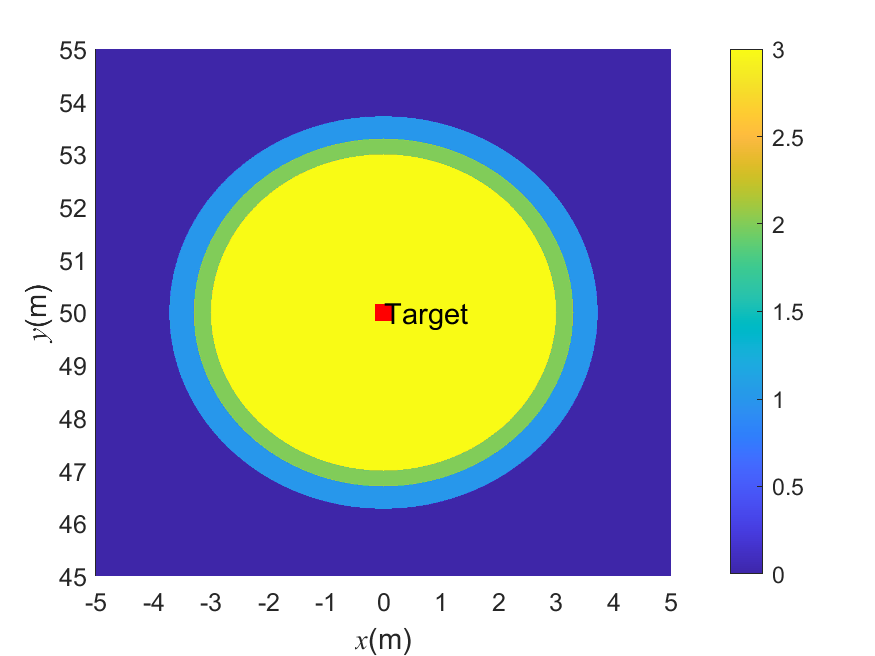}&
			\includegraphics[width=0.53\linewidth]{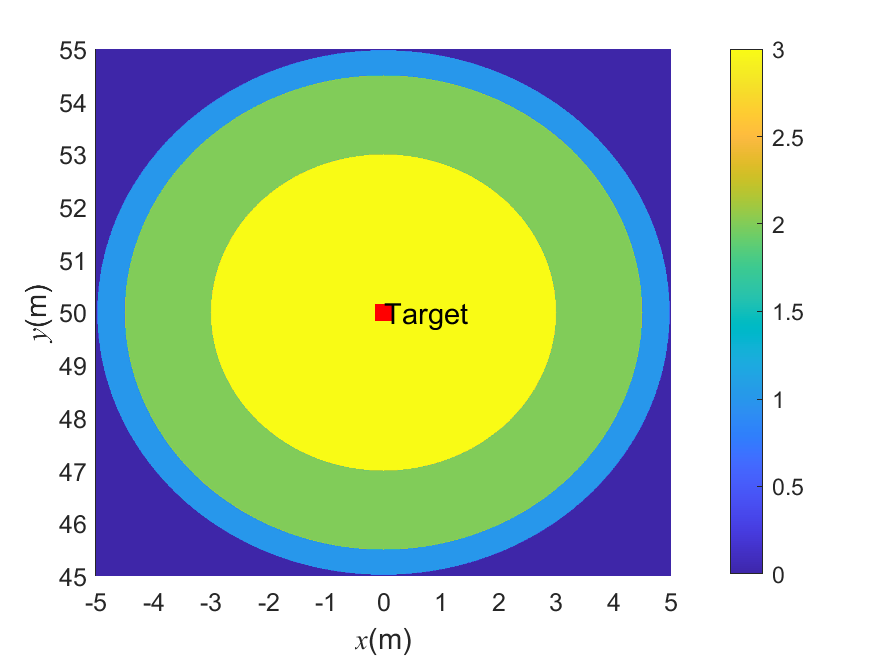}\\
			(a) $P_{md}=0.80$ & (b) $P_{md}=0.50$\\
		\end{tabular}
	\end{minipage}
	\caption{Illustration of spatial registration.}\label{PD}
\end{figure}

Finally, we discuss the effect of spatial registration matching degree\cite{9800700} on performance. In Fig. \ref{tw}, as the matching degree increases, the total network energy consumption decreases, the average echo SINR rises the offloading sum-rate rises, and the execute data rises. The above phenomena are more pronounced as the number of TRIS elements increases, confirming the importance and effectiveness of TRIS and spatial registration for network performance improvement. In addition, a schematic diagram of spatial registration is given in Fig. \ref{PD}, which shows that the higher the matching degree the higher the overlap of the sensing regions.

\section{Conclusions}
In this paper, we investigate a novel TRIS transceiver-empowered ISCC network to meet the diverse needs of future networks. In it, a two-layer computational architecture is designed, where the acquired sensing data is divided into local and remote computation, and the remote computation is offloaded to the MEC through communication.The network considers the cooperative mode, and therefore a spatial registration based optimization algorithm is designed to achieve the cooperative gain, which is implemented through the framework of BCD algorithms. The results show that the TRIS transceiver is able to achieve an overall improvement in the network performance in the form of low energy consumption, which is superior to the traditional transceiver and the baseline scheme, and the cooperative gain is highly affected by the spatial registration matching degree. In addition, the network performance can be improved by increasing the number and power of TRIS elements, which provides guidance for future network design.

\bibliographystyle{IEEEtran}
\bibliography{IEEEabrv,reference}

%
%
%
%
%
%
%
%

\end{document}